\documentstyle[12pt,aasms4]{article} 

\slugcomment{Submitted to the Astrophysical Journal, Part I}

\newcommand{\kms}{{\,\rm km\,s}^{-1}} 
\newcommand{\ksm}{{\,\rm km}\ {\rm s}^{-1}\ {\rm Mpc}^{-1}} 
\lefthead{Saha {\rm et~al.}}
\righthead{SN~1998aq in NGC~3982}

\def\placetable#1{\vspace{0.5ex}\begin{center}EDITOR: PLACE TABLE \ref{#1} HERE.
\end{center}\vspace{0.5ex}}
\def\placefigure#1{\vspace{0.5ex}\begin{center}EDITOR: PLACE FIGURE \ref{#1}
HERE. \end{center}\vspace{0.5ex}}
\def\kms{\nobreak\mbox{$\;$km\,s$^{-1}$}}
\def\fm{\hbox{$.\!\!^{\rm m}$}}
\def\la{\mathrel{\hbox{\rlap{\hbox{\lower4pt\hbox{$\sim$}}}\hbox{$<$}}}}
\def\ga{\mathrel{\hbox{\rlap{\hbox{\lower4pt\hbox{$\sim$}}}\hbox{$>$}}}}
\makeatletter
\renewcommand{\fps@figure}{htbp}         
\renewcommand{\fps@table}{htbp}
\makeatother

\begin{document}

\title{Cepheid Calibration of the Peak Brightness of SNe~Ia. \\ XI.
       SN~1998aq in NGC~3982 \\}

\author{A. Saha} 
\affil{National Optical Astronomy Observatories \\
950 North Cherry Ave., Tucson, AZ 85726} 
\author{Allan Sandage}
\affil{The Observatories of the Carnegie Institution of Washington \\
813 Santa Barbara Street, Pasadena, CA 91101} 
\author{G. A. Tammann} 
\affil{Astronomisches Institut der Universit\"at Basel \\ 
Venusstrasse 7, CH-4102 Binningen, Switzerland} 
\author{A. E. Dolphin, J. Christensen}
\affil{National Optical Astronomy Observatories \\
950 North Cherry Ave., Tucson, AZ 85726}
\and
\author{ N. Panagia\altaffilmark{1}, F.D. Macchetto\altaffilmark{1}}
\affil{Space Telescope Science Institute \\ 3700 San Martin Drive,
Baltimore, MD 21218}

\altaffiltext{1}{Affiliated to the Astrophysics Division, Space
                 Sciences Department of ESA.}

\begin{abstract}
Repeated imaging observations have been made of NGC~3982 with the {\it
Hubble Space Telescope} between March and May 2000, over
an interval of 53 days.  Images were obtained on 12 epochs in the
$F555W$ band and on five epochs in the $F814W$ band.  The galaxy
hosted the type Ia supernova SN1998aq.

A total of 26 Cepheid candidates were identified, with periods ranging from 
10 to 45 days, using photometry with the DoPHOT program.  
The de-reddened distance to NGC~3982 is estimated from these data using 
various criteria to maximize signal to noise and reliability: the values 
lie between 31.71 and 31.82, with uncertainties in the mean of typically 
$\pm 0.14$ mag for each case. A parallel analysis using photometry with 
HSTphot discovered 13 variables, yielding a distance modulus of 
$ 31.85 \pm 0.16 $. The final adopted modulus is $(M-m)_0 = 31.72 \pm 0.14$
($22 \pm 1.5$ Mpc). 

Photometry of 1998aq that is available in the literature is used in 
combination with the derived distance to NGC~3982 to obtain the peak absolute 
magnitude of this supernova. The lower limit (no extinction within the 
host galaxy) for $M_{V}$ is $ -19.47 \pm 0.15 $ mag. Corrections for 
decline rate and intrinsic color to carry these to the reduced 
system of Parodi et~al. (2000) have been performed. 
The derived luminosities at hand are fully consistent with the mean of the 
8 normal SNe~Ia previously calibrated with Cepheids. Together they yield 
$ H_{0} \approx 60 \pm 2 (\rm internal) \ksm $ based on an 
assumed LMC distance modulus of 18.50. We point out that correcting 
some of the systematic errors and including uncertainty estimates due 
to them leads to  $ H_{0} = 58.7 \pm 6.3 (\rm internal) \ksm $.

\end{abstract}

\keywords{Cepheids --- distance scale --- galaxies: individual
(NGC~3982) --- supernovae: individual (SN~1998aq)} 

\section{Introduction} 

This is the eleventh paper of a series whose purpose is to obtain Cepheid 
distances to galaxies that have produced supernovae of type Ia (SNe~Ia), 
thereby calibrating their absolute magnitudes at maximum light. 
The Hubble diagram for SNe~Ia shows a dispersion (read as magnitude residuals 
about a line of slope $dM/d \log cz = 5$) of less than 0.2 mag when certain 
second-parameter corrections are applied (Hamuy et al. 1996a, 1996b; 
Tripp 1998; Phillips et al. 1999; Saha et al. 1999; Tripp \& Branch 1999; 
Parodi et al. 2000; Tammann, Sandage \& Saha 2001), the far-field value of 
the Hubble constant is determined directly from the SNe~Ia Hubble diagram 
once the mean absolute magnitude of SNe~Ia at maximum light is calibrated. 
Clearly the resulting Hubble constant is the global value, free from all 
local velocity anomalies. 


This route to $H_{0}$ through Branch-normal SNeIa 
(Branch et al. 1993; Branch 2001)
is the only method that in a {\it single step} directly bridges the 
relatively nearby distances from  
Cepheids with distances well beyond those corresponding to recession 
velocities of $10,000 \kms$. Such large distances are required 
for the determination of the {\it cosmic} value of $H_{0}$, so that the effect 
of streaming motions is minimized. 
Magnitude residuals of less than 0.2 mag rms in the Hubble diagram show that 
Branch-normal SNeIa, once they are 
corrected for variations in decline rate and intrinsic color, are the 
best known standard candles.
However, before the advent of the {\it Hubble Space 
Telescope (HST)}, one had to rely on other secondary distance indicators to
obtain the luminosity calibration of SNeIa (Sandage \& Tammann 1982), since
the nearest galaxies that had hosted recorded SNeIa were beyond the range 
of ground based Cepheid searches. 
With the use of {\it HST}, this  step of using intermediate secondary 
distance indicators has been eliminated. 
As of this eleventh paper in our series, the peak luminosities of nine 
Branch-normal 
SNeIa are now directly tied to Cepheids. For these reasons 
we claim that the resulting value of $H_{0}$ is the cleanest and most direct
of all the ones using the distance ladder methodology.

The Branch normal type Ia SN 1998aq was discovered by M. Armstrong 
(Hurst 1998) on 1998 April 13, six days before maximum light (see \S 5). 
It appeared 18'' west and 7'' north of the center of NGC 3982, which is a 
galaxy in the very busy region of the supergalactic plane in what was 
originally called the Ursa Major cloud (Humason et al. 1956, Table~I). 
The region was later mapped into separate groups by many cartographers. The 
most complete mapping has been done by Tully (1987), who is the leading 
student of groups. He also summarized much of the earlier work, principally 
by de Vaucouleurs (1975). Later work by Nolthenius (1993) on groups is also 
particularly important to cite. 

What is now called the Ursa Major Cluster is in the midst of the Ursa Major 
Cloud (see map 15 of the Tully-Fisher Atlas, 1987). The most convincing 
mapping of the region, and the separation of the Ursa Major Cluster from the 
Cloud, is by Tully et al. (1996). following the thesis of Verheijen (1997). 
NGC 3982 lies near the extreme northern border of the Cluster as delineated by 
Verheijen and by Tully et al. (1996). The galaxy is listed in the RSA (Sandage 
\& Tammann 1987) as of type Sbc(r)II-III with a total blue apparent magnitude, 
corrected for Galactic and internal absorption, of $B_{T}^{b,i} = 11.59$. 
Tully lists the apparent magnitude, similarly corrected by his absorption 
precepts, as $B_{T}^{b,i} = 11.7$. A color image of this galaxy, created 
from the the HST $V$ and $I$ band images presented in this paper, is shown 
in Fig.~\ref{fig1}. 

Membership in the the Ursa Major Cluster is of no importance in 
calibrating the absolute magnitude at maximum of its daughter supernova, 
which is the object of this paper. Nevertheless, our determination of the 
distance to the galaxy in \S 4, does, of course, 
have relevance for comparing the distance to the Ursa Major Cluster 
summarized by Tully et al. (1996) to be 15.5 Mpc from other methods. This 
distance, along with their quoted distances of 15.6 Mpc for the Virgo 
cluster and 14.5 Mpc for the Fornax cluster, defines their short distance 
scale leading to their high value of $H_{0} = 80$ (Pierce \& Tully 1988). 
Our distance to NGC 3982 from Cepheids (\S 4 of this paper) is 22.1 Mpc.
Combined with the apparent magnitude at maximum of SN 1998aq and the absolute 
magnitudes at maximum light from eight previous SNe~Ia calibrators 
(Saha et al. 1999; Parodi et al. 2000) leads us again to 
$H_{0} = 59 \pm 6 {\rm km s^{-1} Mpc^{-1}}$.

The plan of this paper is as follows. The HST observations of NGC~3982
are set out in \S 2. Discovery and photometry of the Cepheids is described 
in \S 3. The P-L relation and the resulting determination of the apparent as
well as absorption-corrected distance moduli are in \S 4. Discussion of the 
available photometric data for SN 1998aq and the resulting calibration of 
its absolute magnitude is in \S 5. In \S 6 we compare $M(max)$ for SN 1998aq 
with the previously calibrated nine SNe~Ia, leading to our determination 
of $H_{0}$ using the supernova route.

\section{Observations}

Repeated images of a field that contains the entire visible region 
of the galaxy NGC~3982 were obtained using the WFPC2 (\cite{hol95a}) on 
the {\it HST} over a 53 day period from 2000 March to May (HST observing 
program number 8100).
The visible galaxy fits within the field of view of the WFPC2, and a 
composite image including all four chips of the instrument is shown 
in Fig.~\ref{fig1}. Observations were made at 12 discrete epochs in 
$F555W$ passband, and at 5 epochs in the $F814W$ passband over the 
53 day window. Each epoch in each filter consists of 2  
one-orbit duration exposures taken on successive orbits of the spacecraft.
This allows the removal of cosmic rays by an anti-coincidence technique
described by \cite{saha96a} (Paper V). The journal of observations is 
given in Table~\ref{tbl1}. All exposures in this table are pairs of 
exposures on back to back space-craft orbits, each of 2500s duration.

\placetable{tbl1}

The epochs were spaced strategically over the total duration to provide 
maximum leverage on detecting and finding periods of Cepheid variables over
the period range from 10 to 50 days. It should be remarked that it is 
highly desirable to keep all the observations at very nearly the same 
pointing as possible. We have managed to do this in the previous 
galaxies studied in this series. Due to a combination of a lack of guide 
stars for this field at some spacecraft roll angles and the pointing 
constraints posed by the requirements for the solar panels, a large 
enough window for keeping the pointing orientation unchanged was not 
available in the present case. Consequently, 
while the final 11 epochs are all obtained with 
essentially the same pointing (to within a few pixels), the very 
first epoch ($F555W$ filter only) has a field orientation that is 
12 degrees different from the remaining observations. 

\section{Photometry}

\subsection{DoPHOT based analysis}

The procedural details for processing images, combining the sub-exposures for 
each epoch while removing the bulk of the cosmic rays, as well as performing 
the photometry with a variant of DoPHOT (\cite{schec93}) optimized for 
WFPC2 data 
has been given in Paper~V. A subsequent improvement in deriving aperture 
corrections 
was detailed in \cite{saha99} (Paper~IX). Since the same procedures were 
followed here, the details are not repeated. 

\subsubsection{A note on magnitude zero points}

In keeping with the precepts mentioned in Paper~V, measurements in any one 
passband are expressed in the magnitude system defined by \cite{hol95b} that 
is natural to the HST. Specifically we mean the $F555W$ and $F814W$ 
``ground system'' magnitudes. These calibrations were made with ``short''
exposures, which have since been shown to suffer from the effects of 
anomalous charge transfer in the CCD devices used (e.g. \cite{whit99}).
It is recognized that the effect of adopting the \cite{hol95b} calibrations
over-estimates the brightness by a few hundredths of a magnitude. In our 
previous papers of this series, we made a correction of 0.05 mag to both
passbands to account for this effect. Our current understanding of this 
situation is that the exact correction is procedure dependent
(\cite{saha00}). In addition, the charge transfer anomaly has been shown to 
worsen with time due to exposure to radiation. Thus the exact corrections 
depend additionally on when a particular data-set was obtained.
While \cite{saha00} evaluated the appropriate corrections for short exposure 
data taken in late 1997, that analysis does not directly predict the 
corrections appropriate for the \cite{hol95b} calibration.
The results of the bright star monitoring program indicate that 
the deterioration from 1994 to 1997 are unlikely to be significant.
Nevertheless, the results of future study (which includes some of the 
authors of this paper) could adjust the photometric zero points at the 
0.03 mag level. 

In this paper we continue to show all $F555W$ and $F814W$ magnitudes on 
the \cite{hol95b} system. Transformed magnitudes in $V$ and $I$ are shown 
with the 0.05 magnitude corrections, which is consistent with the final 
numbers shown in the previous papers of this series. The data can be updated 
for better corrections as they become available. 

\subsubsection{Discovery and Classification of the Variable Stars}

The measured magnitudes and reported errors at all available epochs in 
$F555W$ were used to identify variable stars using the method described
by \cite{saha90}. The procedural details specific to the WFPC2 data in 
this series of papers has been given in Paper~V, \cite{saha96b} (Paper~VI)
and \cite{saha97} (Paper~VIII). They are not repeated here. 

All variable stars that were definitely so identified, are marked in 
Fig.~\ref{fig2}. However some of the variables are not visible on these 
charts due to their extreme faintness combined with the variation in the 
background surface brightness on these images. The positions of these 
objects, as they appear in the images identified in the HST archives as 
$U5KY0201R$ and $U2KY0202R$  are listed in Table~\ref{tbl2}. 

\placefigure{fig2}     \placetable{tbl2}

The photometry on the \cite{hol95b} ``short exposure'' calibration
system for the final list of 26 variable stars is presented in
Table~\ref{tbl3} for each epoch and each filter.  The periods were
determined with the \cite{laf65} by using only the $F555W$ passband
data.  Aliasing is not a serious problem for periods between 10 and
55 days because the observing strategy incorporated an optimum
timing scheme as before in this series. 

\placetable{tbl3}

\begin{figure}
\plotone{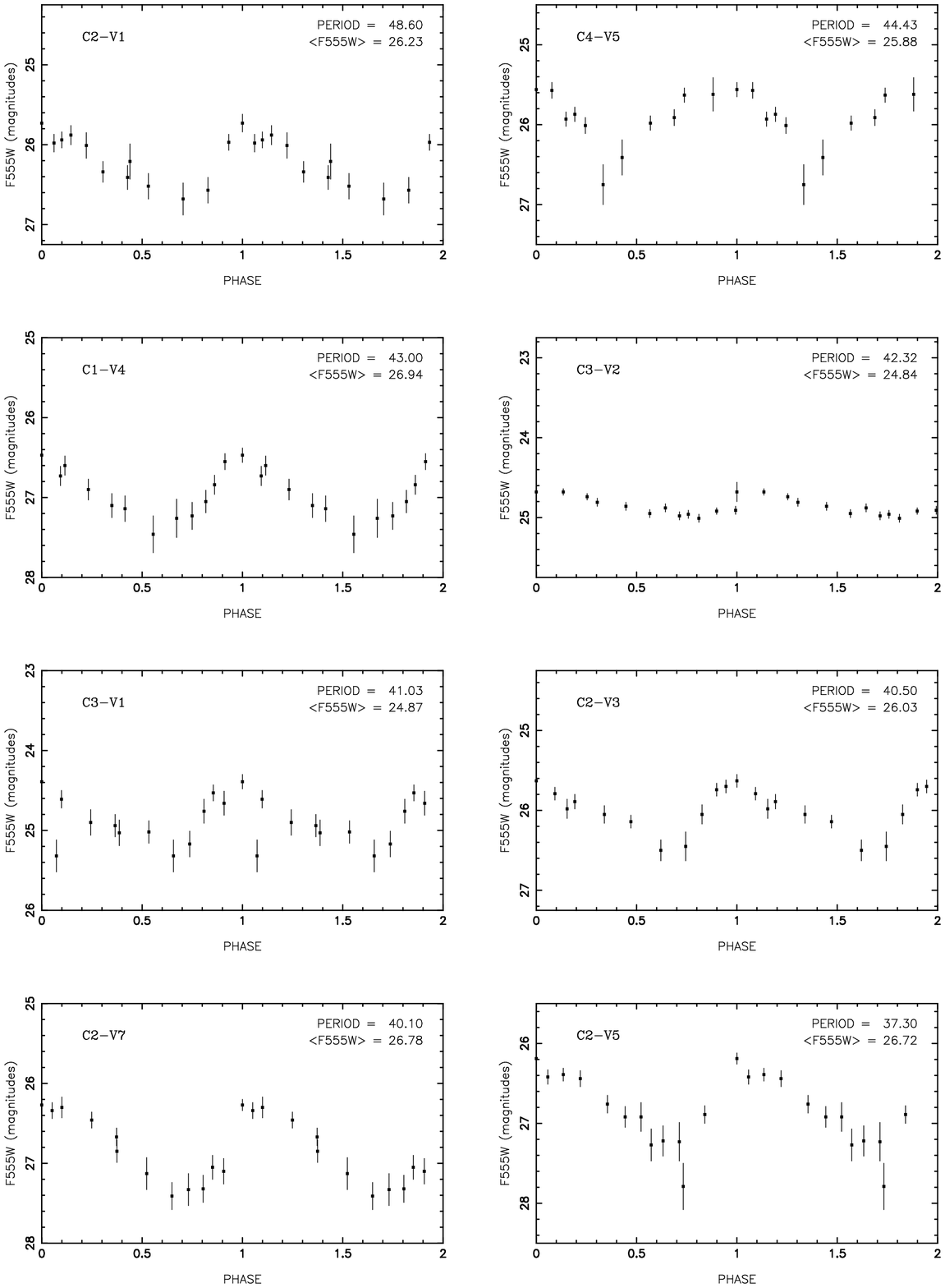}
\figcaption[]{Light curves, plotted in order of period, in the $F555W$
band. \label{fig3}}
\end{figure}

The resulting light curves in the $F555W$ passband, together with
periods and mean magnitudes (determined by integrating the light
curves, converted to intensities, and then converting the average
back to magnitudes, and called the ``phase-weighted intensity average'' in
\cite{saha90}), are shown in Fig.~\ref{fig3}, plotted in the order of
descending period. 

\placefigure{fig3}

All of the variable stars identified in this manner have  
periods and light curves consistent with being Cepheids. 
There is a range in the quality of the light curves -- both in terms 
of the scatter in the individual points, as well as in the implied 
shapes. The object C3-V2, for instance, has an extremely small amplitude, 
even though its variability appears quite definite in signal-to-noise 
terms. It is possible that this is a blend of two stars, one of which 
may be a Cepheid. Thus sorting through the quality of each putative Cepheid
is a necessary task, and is described later in this sub-section. 

\begin{figure}
\plotone{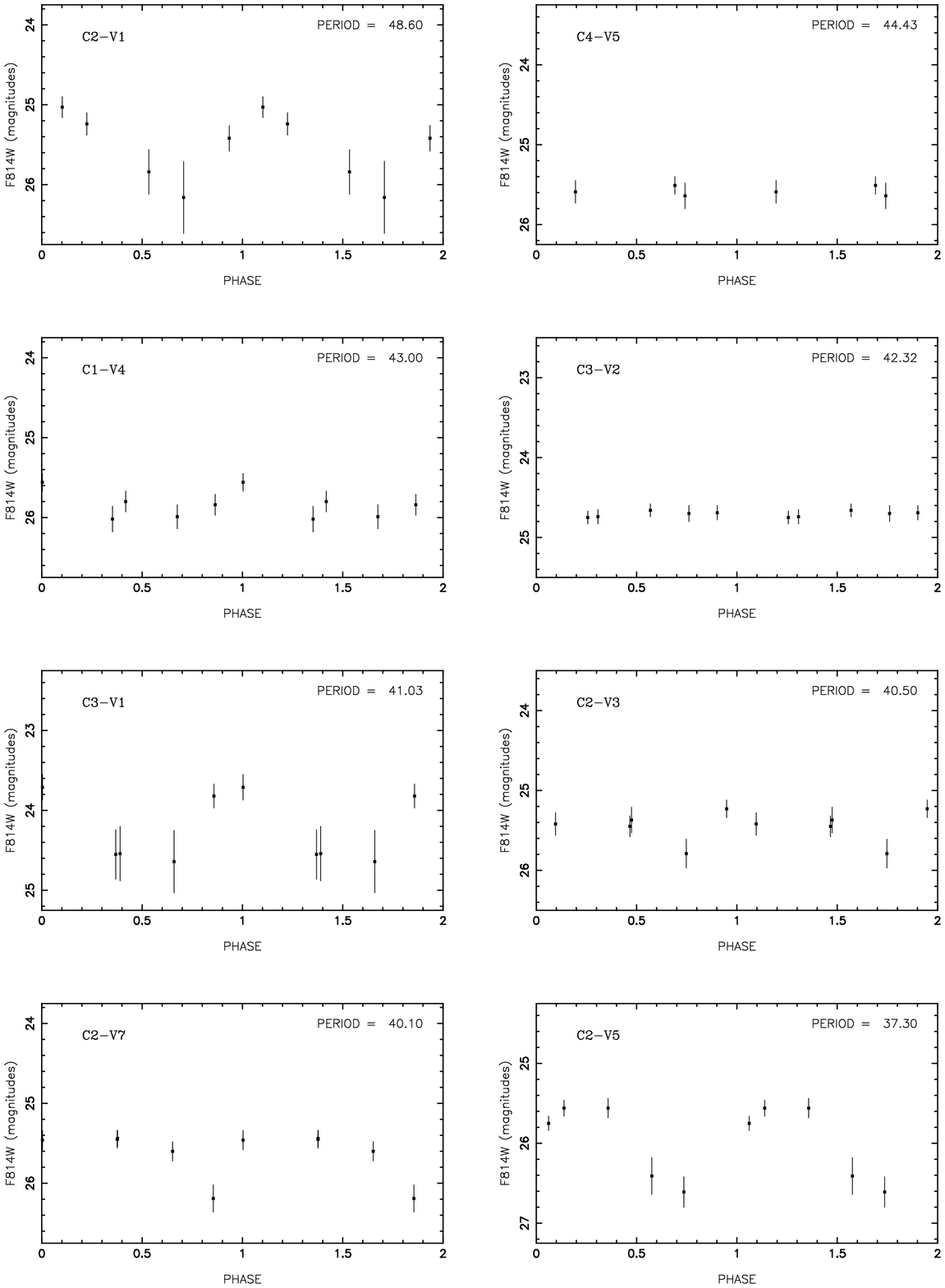}
\figcaption[]{Same as Fig.~\ref{fig3} but for the $F814W$
passband, adopting the periods and the phasing used in
Fig.~\ref{fig3}. \label{fig4}}
\end{figure}

The available data for the variables in $F814W$ were folded with the
ephemerides derived above using the $F555W$ data.  The results are
plotted in Fig.~\ref{fig4}. All of the objects detected as variables in $F555W$
were recovered in at least 3 epochs in the $F814W$ observations.  

The mean magnitudes in $F814W$ (integrated as intensities over the
cycle) were obtained from the procedure of \cite{lab97} whereby each
$F814W$ magnitude at a randomly sampled phase is converted to a mean value
$\langle{F814W}\rangle$ using amplitude and phase information from
the more complete $F555W$ light curves.  Note that each available
observation of $F814W$ is used independently to derive a mean
magnitude.  Hence, the scatter of the individual values about the
adopted mean $F814W$ value is an {\it external} measure of the
uncertainty in determining $\langle{F814W}\rangle$.  It is this
external measure of the uncertainty that is retained and propagated in
the later calculations. 

\placefigure{fig4}

The prescription given in Paper~V for assigning the light-curve
quality index $QI$ (that ranges from 0 to 6) was used.  In this scheme,
a grade ($0-2$) is given for the quality of the $F555W$ light curves, 
with additional points ($0-2$) for the evenness in phase coverage of the 
five (or fewer) $F814W$ observation epochs, and up to three additional 
points ($1-3$) for the amplitude and phase
coherence of the $F814W$ observations compared with the $F555W$ light
curve.  Hence, a quality index of 6 indicates the best possible light
curve set in both $F555W$ and $F814W$. 
A quality index of 2 or less indicates near fatal flaws
such as apparent phase incoherence in the two passbands.  This is
generally the indication that object confusion by crowding and/or
contamination by background is likely. The weighting scheme puts a lot 
of weight on how well matched the few $F814W$ observations are to the 
light curve implied by the $F555W$ data. 
This is by design, since if 
reddening effects are to be deciphered from the Cepheid colors, the 
fidelity of the colors must be established, since color uncertainties 
dominate the error in the de-reddened distance modulus.

Table~\ref{tbl4} lists the characteristics of all 26 putative Cepheids 
mentioned above. The
$F555W$ and $F814W$ instrumental magnitudes of Table~\ref{tbl3} have
been converted to the Johnson $V$ and Cousins (Cape) $I$ standard
photometric system by the color equations used in previous papers of
this series, as set out in equations (2) and (3) of Paper~V, based on
the transformations of \cite{hol95b}. In addition, as discussed in the 
previous sub-section, a value of 0.05 mag has been added to each of the 
$V$ and $I$ magnitudes to correct the \cite{hol95b} scales for 
the charge-transfer inefficiency problems. This correction is consistent 
with that used in all previous papers of this series, except that in the
previous papers, the correction was applied at the end, to the distance 
moduli, and not in the tables corresponding to Table~\ref{tbl4} here.

\placetable{tbl4}

\subsection{HSTphot based photometry}

A parallel photometry procedure was carried through by one of us (AED)
using the HSTphot stellar photometry package \cite{dol00a}.  The images
were masked using the data quality images, and pairs of images combined
for cosmic ray removal using HSTphot's \textit{crclean} algorithm,
producing eleven F555W and five F814W images (all 5000s) at the primary
pointing, and a twelfth F555W image (also 5000s) at the secondary
pointing.

The \textit{multiphot} algorithm was used to simultaneously photometer
all sixteen images at the primary pointing.  Because of the distance to
NGC 3982, it was extremely difficult to locate individual stars for use
as PSF stars.  Instead, the photometry was run twice: the first time
using a library PSF calculated from Tiny Tim (\cite{kri95}) models, and
the second time adding a residual (calculated from the stars found in
the first run) to the PSF.

This process was also completed on the single image at the secondary
pointing.  To arrive at our final HSTphot photometry, we used the
\cite{dol00b} formulae to make CTE corrections and calibrate to the
standard $VI$ system, and matched the secondary pointing photometry to
the primary pointing, producing instrumental and standard magnitudes for
each star at each epoch.


The post-photometry procedure to detect and characterize the Cepheids
was essentially identical to that used with the DoPHOT photometry 
results.  The main differences involved the minimum acceptable photometry
quality used -- the HSTphot analysis required that stars be found with
$\chi < 2.0$ and $|\mbox{sharpness}| < 0.4$ in order to avoid blends and
stars for which good photometry was impossible.  We also required that
a star have at least 10 such good photometry measurements to be considered
in the analysis.  The quality parameters in the HSTphot-based analysis
range from 0-4 (rather than 0-6), but are based on similar criteria --
cleanness of both the F555W and F814W light curves, coherence between the
light curves, and phase coverage.

Of the 26 variables found by DoPHOT, 13 were independently discovered 
from the HSTphot analysis. Ten of these 13 common Cepheids have HSTphot 
quality parameters of 3 or 4.
The periods and mean $V$ and $I$ magnitudes derived from 
the HSTphot measurements alone for these 13 Cepheids are given in 
Table~\ref{tbl5}. 

The results from the HSTphot based procedure are used to estimate the 
distance modulus in \S4.3, and used as a sanity check for the 
results from DoPHOT. 

\placetable{tbl5}

\section{The Period-Luminosity Relation and the Distance Modulus}

\subsection{ The P-L Diagrams in $V$ and $I$ }

As in the previous papers of this series we adopt the P-L relation in
$V$ from Madore \& Freedman (1991) as 

\begin{equation}
 M_{V} ~~=~~ -2.76 ~\log P - 1.40~, 
\end{equation} 
whose companion relation in $I$ is 
\begin{equation}
 M_{I} ~~=~~ -3.06 ~\log P - 1.81~.  
\end{equation} 
The zero point of equations (1) and (2) is based on an assumed LMC
modulus of 18.50.

\begin{figure}
\plotone{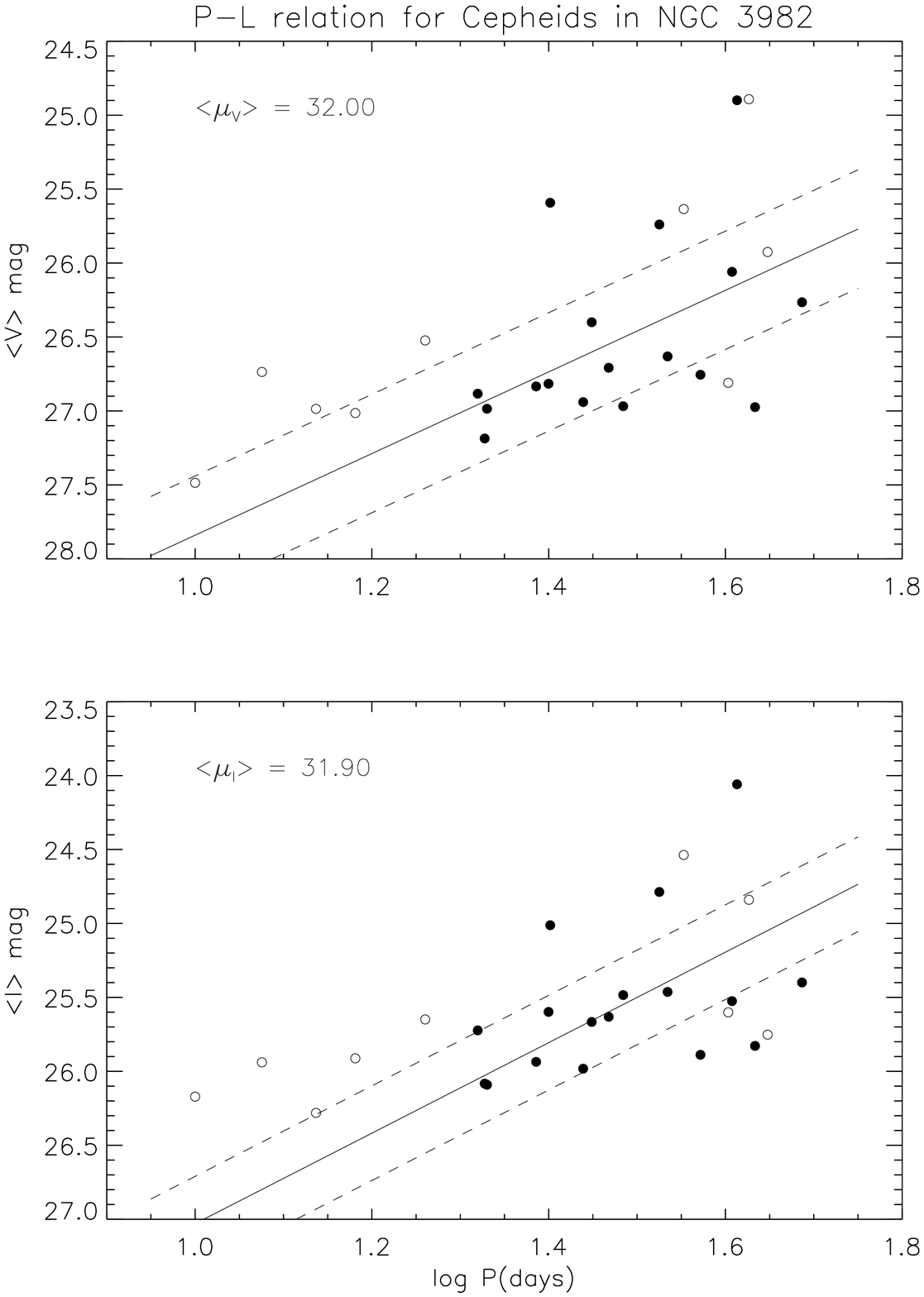}
\figcaption[]{The apparent P-L relations in $V$ and $I$ showing
all the Cepheid data in Table~\ref{tbl4}.  The solid circles are
Cepheids with $ P \geq  20^{d}$ with quality index $QI$ of 3 or better.  
Open circles are for the remainder of the variables in
Table~\ref{tbl4}. The drawn lines are the adopted P-L relations from
Madore \& Freedman (1991) in
equations (1) and (2) used also in the previous papers of this series.  
The ridge lines have been put arbitrarily at a 
modulus of
$(m -M)_{V} = 32.00$ and $(m-M)_{I} = 31.90$ as a first estimate, 
before analysis of differential extinction. The dashed upper and lower 
parallel lines indicate the expected projection of the width of the 
instability strip. The heavy spillage of points outside these bounds indicates
the presence of large differential extinction or noise or both. 
\label{figPLR1}}
\end{figure}

The P-L relations in $V$ and $I$ for the 26 Cepheids in
Table~\ref{tbl4} are shown in Fig.~\ref{figPLR1}.  The filled circles
show objects with periods greater than 20 days that have a quality
index of 3 or higher. 
The solid lines show the canonical slopes 
of the P-L relations in $V$ and $I$ with the vertical offset 
for apparent distance moduli 
$\mu_{V} = 32.00$ and $\mu_{I} = 31.90$ respectively. 
These values were chosen  
to be in visual conformity with the points shown as filled circles
(they are not intended as a formal derivation of distance). 
The expected spread in each of the pass-bands due to the 
finite width of the instability strip (\cite{sata68}) is indicated by 
the flanking dashed lines.
The observed scatter of the data outside these envelope lines
is due to the combination of (1) measuring and systematic errors due
to background and contamination, (2) the random error of photon
statistics, (3) the large effects of the variable extinction
evident from the dust lanes seen in the images, and (4) objects mis-identified
as Cepheids.

\placefigure{figPLR1}

The variables with periods shorter 
than 20 days can be seen to fall systematically brighter. This is due to a  
bias at the faint end because Cepheids at short periods that are at the 
faint end of the intrinsic scatter about the mean P-L relation and fall below 
the detection limit in brightness do not populate the P-L relation. This 
adversely affects the fitting of the P-L relation. A cut off that rejects 
objects shorter than 20 days is a sensible precaution in this case, and such 
a period cut does not introduce a bias of its own. 

\subsection{Analysis of the P-L Relation} 

For a first estimate, using $A_{V}/A_{I} = 1.7$
(\cite{schef82}) along with the very preliminary 
apparent moduli in $V$ and $I$ of 
32.00 and 31.90 respectively as estimated 
above yields a dereddened modulus $\mu_{0} \approx 31.76$.
To explore the presence of differential extinction and 
to treat the data accordingly, we use the tools developed in Paper~V and 
further developed in subsequent papers of this series.

For each Cepheid we calculate the apparent distance moduli separately
in $V$ and in $I$ from the P-L relations of equations (1) and (2) and
the observed $\langle{V}\rangle$ and $\langle{I}\rangle$ magnitudes
from Table~\ref{tbl4}. These apparent distance moduli, called $U_{V}$
and $U_{I}$ in columns (7) and (8) of Table~\ref{tbl4}, are calculated
by
\begin{equation}
 U_{V} ~~=~~ 2.76 ~\log P + 1.40 + \langle{V}\rangle~, 
\end{equation}
and 
\begin{equation}
 U_{I} ~~=~~ 3.06 \log P + 1.81 +\langle{I}\rangle~.  
\end{equation} 
They are the same as equations (6) and (7) of Paper~V.

If the differences between the $V$ and $I$ moduli are due solely to
reddening, and if the dependence of the reddening curve on wavelength
is the normal standard dependence as in the Galaxy, then the true
modulus $U_T$ is given by 
\begin{equation}
 U_{T} ~~=~~ U_{V} - R'_{V} \cdot (U_{V} - U_{I})~, 
\end{equation} 
where $R'$ is the ratio of total to selective absorption, $A_{V}/E(V-I)$.
This is equation (8) of Paper~V. However, equation (5) is valid only
if the difference between $U_{V}$ and $U_{I}$ is due to extinction,
not to correlated measuring errors. 

The values of $U_{T}$ are listed in column 9 of Table~\ref{tbl4}.
These would be the true moduli, as corrected for normal extinction,
assuming that there are no systematic measuring errors.  The total rms
uncertainty for each $U_{T}$ value is listed in column 10.  This
uncertainty includes contributions from the estimated random measuring
errors in the mean $V$ and $I$ magnitudes, (in columns 4 and
6), as propagated through the de-reddening procedure, as well as the
uncertainty associated with the intrinsic width of the P-L relation
(i.e. a given Cepheid may not be on the mean ridge-line of the P-L
relation) as well as a ten percent uncertainty in the estimated period.
 The de-reddening procedure amplifies the measuring
errors. Therefore many Cepheids are needed to beat down these large
errors (notice some very large values in column 10) in any final value
of the modulus.  The values shown in column 10 of Table~4 were calculated 
using equation (18) of \cite{saha00}, and also correspond to $\sigma_{tot}$
as defined in Paper~V.  

From the data in Table~\ref{tbl4}, the unweighted mean de-reddened modulus 
$\mu_{0} = <U_{T}>$ for  all Cepheids 
with periods greater than  20 days with $QI \geq 3$ is $ 31.72 \pm 0.14 $ mag, 
and the weighted (by $1/\sigma_{tot}^2)$ value is also $ 31.72 \pm 0.15 $ mag. 
The average (unweighted) {\it apparent} modulus $\mu_{V}$ for 
this sample of Cepheids is 
$ 31.99 \pm 0.13 $ and the 
corresponding value for 
$\mu_{I}$ is $31.88 \pm 0.12$. This implies $E(V-I) = 0.11 \pm 0.17$ and 
$A_{V} = 0.27 \pm 0.41$, 
where the symbols have their usual meaning.

We mention again that the derived $U_{T}$ values are only meaningful under 
the assumption that the differences between $U_{V}$ and $U_{I}$ are due to
reddening alone, in the absence of appreciable systematic and
correlated measuring errors, or when the errors for $U_{V} - U_{I}$ {\it
are distributed symmetrically}.  
If asymmetrical errors in $V$ and
$I$ dominate over differential reddening,
the $U_{T}$ derived via equation (5) {\it will be systematically 
in error}.

\begin{figure}
\plotone{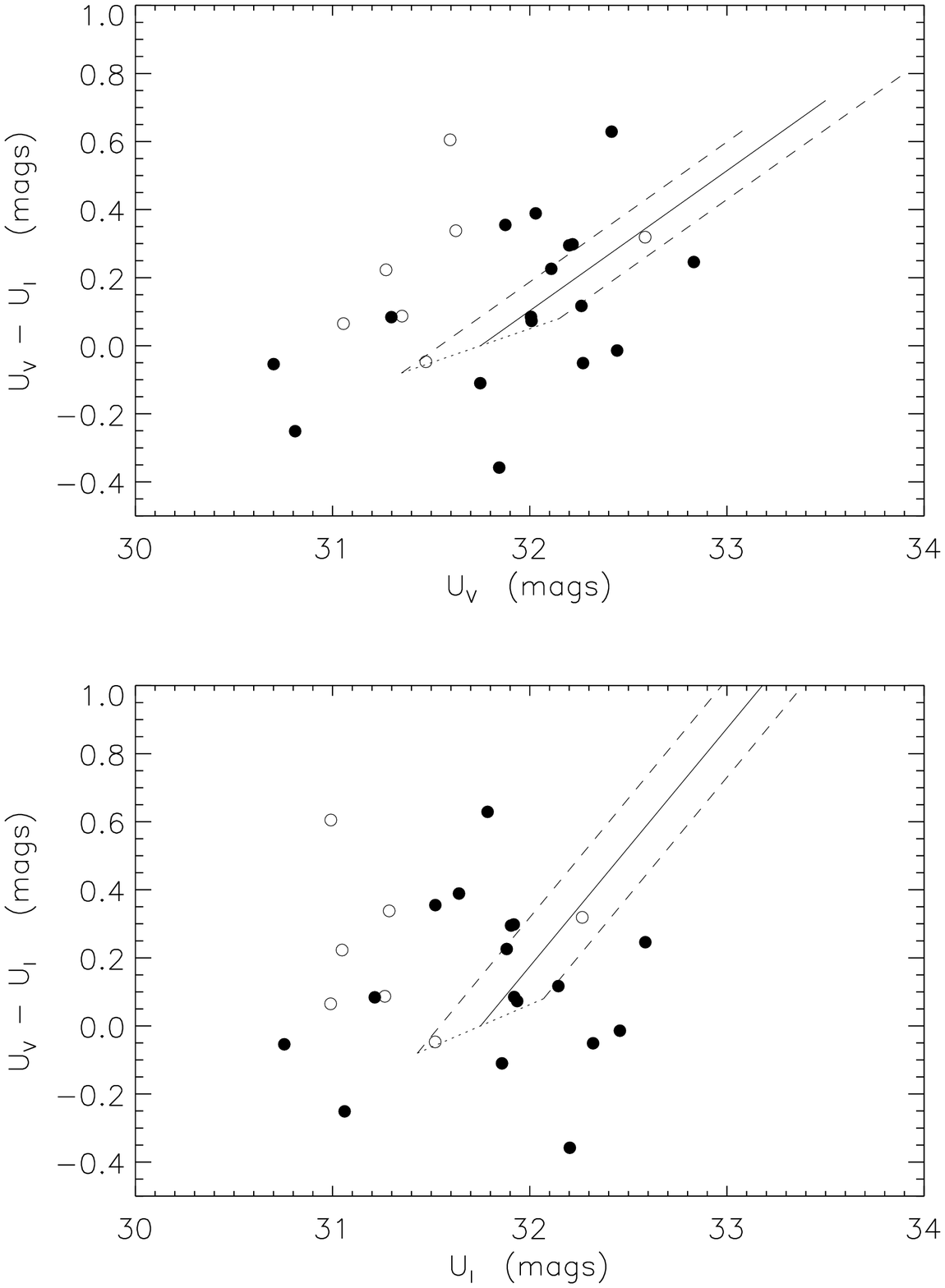}
\figcaption[]{Diagnostic diagram for the detection of differential
reddening, described in the text. The top panel shows the $V$-modulus 
on the abscissa. The lower panel is the same figure, but with 
the $I$-modulus on the abscissa. 
\label{figDIAG}}
\end{figure}

A diagnostic diagram to test for the relative presence of bona-fide 
differential extinction versus scatter due to measuring errors alone was
devised in Paper~V, and used in subsequent papers of this series. We do not 
repeat its description. It is shown in Figure~\ref{figDIAG} for the 
Cepheids in NGC~3982. The filled circles again show Cepheids with periods
greater than 20 days and with $QI \geq 3$. The solid line indicates the 
reddening vector for the P-L ridge
line, if the true (de-reddened) distance modulus is $31.75$ which is close to 
the initial estimate for the distance modulus made above. The dashed lines
show the bounds due to the intrinsic dispersion of the P-L relation as
explained in Paper~V.  The slope of the lines is $A_{V}/E(V-I) = 2.43$, 
in accordance with the reddening law of Scheffler (1982).  

\placefigure{figDIAG}

The scatter in Fig.~\ref{figDIAG} does not lie within the reddening track.
Translating the reddening band in this 
figure would not materially reduce the number of points that would spill 
out of it. There may be some differential reddening, but there is very 
significant scatter orthogonal to the reddening track, indicating that the 
range in $U_{V}$ and $U_{I}$ values is driven more by photometry 
errors rather than by extinction alone. As mentioned above, and 
discussed in previous papers of 
this series, any skewness in the distribution of measurement 
errors contributes to a systematic error in the de-reddened modulus for 
the galaxy. 

\subsubsection{Distance Modulus from a Color-Selected Sample}

Since we have shown that the scatter in observed colors of the Cepheids
is significantly due to observational errors, we can try to exclude 
color outliers by selecting out objects whose measured colors are extreme
for Cepheids. To do this, we utilize the Period-Color (P-C) relation 
for Cepheids, which follows immediately from equations (1) and (2):
\begin{equation}\label{PC1}
   (V-I)_{0} = 0.30 \log P + 0.41
\end{equation}
where $(V-I)_{0}$ is the intrinsic color of a Cepheid on the ridge line 
of the P-L relation. Note that the intrinsic scatter in color is small: the 
delimiting lines shown in Fig~\ref{figPLR1} map to lines 0.08 mag 
above and below the line defined above by eqn. (\ref{PC1}).  
The ridge line summarized in \cite{sanetal99} from independent data 
by \cite{dea78}, \cite{cal85} \& \cite{fern90} is essentially identical 
to the above, as shown in \cite{saha01} (Paper~X). 
The observed colors can be 
reddened by dust, and scatter will be introduced both, by 
differential extinction as well as measurement errors. 

We expect to see some Cepheids that are reddened very little because they 
are on the near side of the disk. Others may have substantial reddening
since they are seen through the disk, and possibly through dust lanes.
The rationale for a color-cut is that while bona-fide extinction should not 
affect the de-reddened distance modulus irrespective of how the data are cut in
color, objects with large errors and objects mis-classified as Cepheids, which 
{\it do} affect the modulus, can be identified and rejected.

\begin{figure}
\plotone{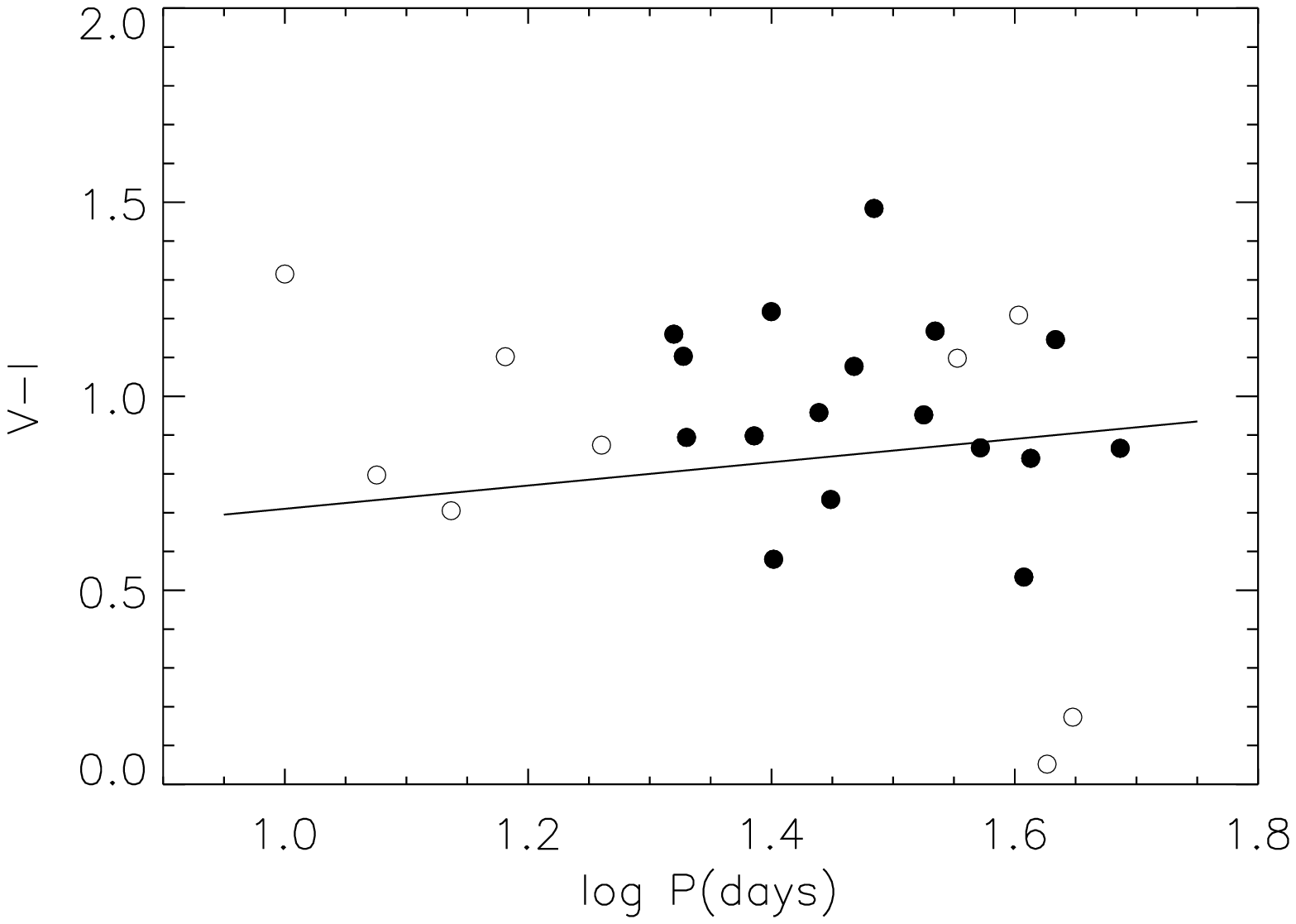}
\figcaption[]{The Period Color Relation in $V$ and $I$. The observed data 
are shown, with filled circles indicating Cepheids with $ P \geq  20^{d}$, 
and $QI \geq 3$. The line shows the fiducial relation for unreddened 
Cepheids that lie on the ridge line of the P-L relation. 
\label{figPCrel}}
\end{figure}

The P-C relation for the data at hand is shown in Fig~\ref{figPCrel}.
As before, the filled circles indicate Cepheids with 
$P \geq 20$ and $QI \geq 3$.

\begin{figure}
\plotone{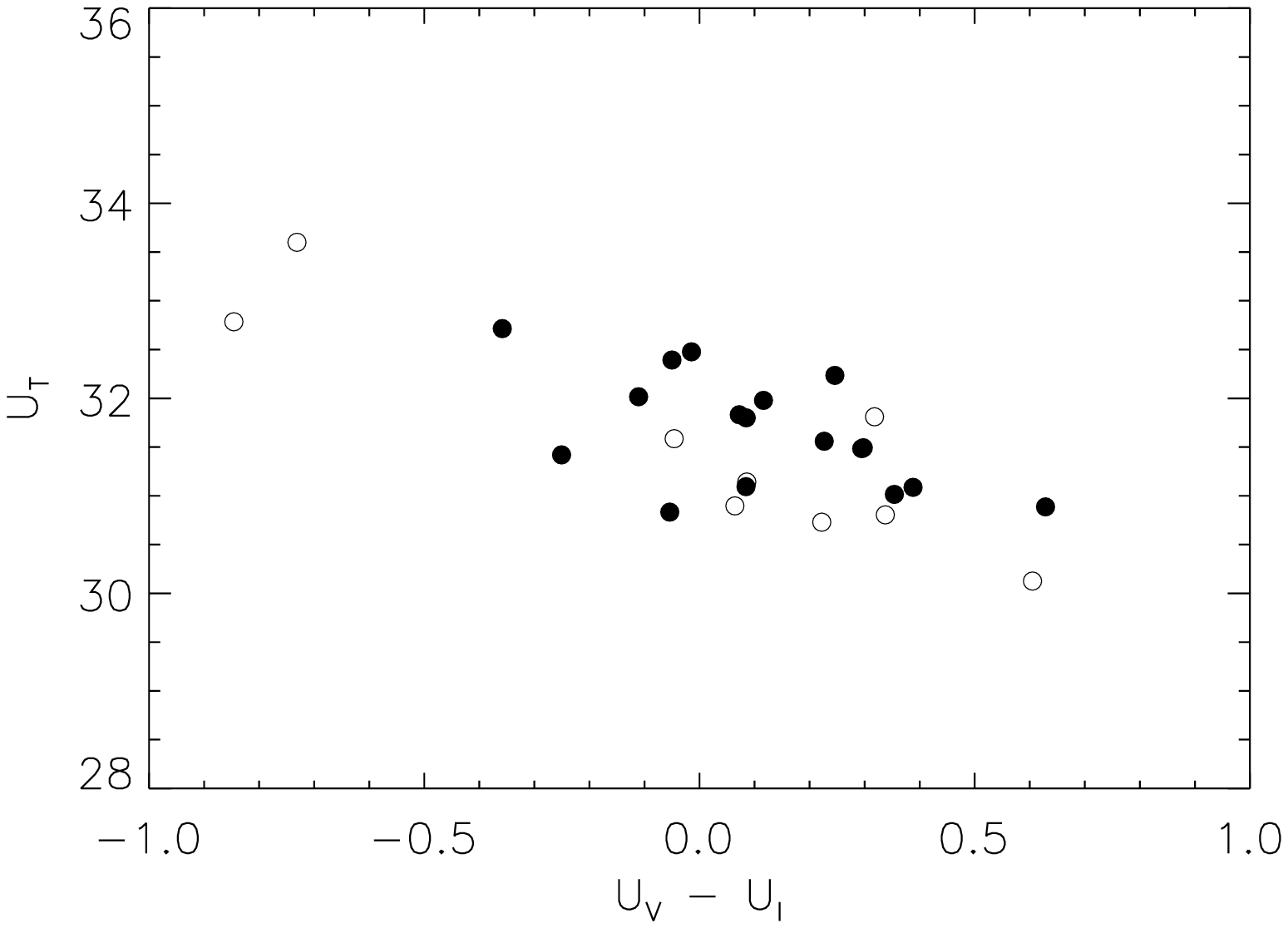}
\figcaption[]{Diagram showing the dependence of implied de-reddened modulus 
as a function of color deviation from the mean P-C relation. The slope is due 
to misinterpretation of measurement errors as reddening. Outliers with 
extreme colors are easily identified as being far from the clump of the 
majority of the objects. Again, filled circles indicate Cepheids 
with $ P \geq  20^{d}$, and $QI \geq 3$.
\label{figPCdev}}
\end{figure}

A plot of individual de-reddened moduli 
$U_{T}$ vs. color deviation 
$\Delta (V-I) ~=~ U_{V} - U_{I}$
from the fiducial PC relation 
for the data at hand is shown in Fig~\ref{figPCdev}, again 
where filled circles indicate Cepheids with 
$P \geq 20$ and $QI \geq 3$. This diagram graphically 
demonstrates that the derived distance is a strong function of the measured 
color, showing again that the data are dominated by measurement errors.
An inspection of this figure indicates that values of $U_{V}-U_{I}$ less 
than $-0.2$ or greater than $0.4$ are almost certainly outliers: 
rejecting these 
and working with the 14 Cepheids with $P \geq 20$ and $QI \geq 3$ yields 
a de-reddened modulus $\mu_{0} = 31.71 \pm 0.14$. This cut corresponds 
to a region delimited by a line 0.2 mag below and a line 0.4 mag 
above the ridge line of the fidicial PC 
relation shown in Fig~\ref{figPCrel}. 

A way to arrive at a less arbitrary cut 
is to try various cuts by color, and choose the one(s) 
that produce the P-L relations with the least scatter in both passbands. 
Following such a path, we arrive at a cut with 
$-0.2 \leq U_{V}-U_{I} \leq 0.3$, which corresponds 
to a line 0.2 mag below and a line 0.3 mag 
above the ridge line of the fiducial PC relation. 
This cut yields $\mu_{0} = 31.82 \pm 0.14$, using the 12 
acceptable Cepheids that also have $P \geq 20$ and $QI \geq 3$.

\subsection{Distance Estimate from the HSTphot results}

We consider here the sample of Cepheids found using the HSTphot photometry, 
as described in \S 3.2. Recall that 13 Cepheids were reported, 
all in common with 
the DoPHOT based discoveries. A different grading scheme with grades from 
0-4 was used, and assigned independently. Using the 10 Cepheids which 
have $P \geq 20^{d}$ and grade 3 or better, and computing the de-reddened 
moduli $U_{T}$ we find that C1-V2 is a clear outlier (too large a modulus).
Using the remaining 9 Cepheids, we arrive at a mean (unweighted) 
de-reddened distance modulus $\mu_{0} = 31.85 \pm 0.16$.  
This is within the errors of the similar calculation for DoPHOT based 
Cepheids. This subsample of 9 Cepheids with HSTphot photometry gives
$\mu_{V} = 32.26 \pm 0.08 $ and $\mu_{I} = 32.09 \pm 0.09$, implying 
$E(V-I) = 0.17 \pm 0.12$ and $A_{V} = 0.41 \pm 0.29$.

\subsection{The Adopted Distance Modulus}

We are satisfied that the results from HSTphot are consistent within 
the errors with the DoPHOT results, and having used this as a sanity check, 
proceed with the DoPHOT values.

The various cuts on the DoPHOT sample of Cepheids were discussed in 
\S 4.2. The results do not change significantly within the errors 
irrespective of the subsample used. We therefore adopt the most robust 
cut of all, which is the one with $P \geq 20^d$ and $QI \geq 3$, which 
was shown to yield:

\begin{equation}\label{DIST1}
  \mu_{0} = 31.72 \pm 0.14
\end{equation}
and 
\begin{equation}\label{REDDEN1}
   E(V-I) = 0.11 \pm 0.17
\end{equation}

The apparent distance modulus in the $V$ band is given by 

\begin{equation}\label{VDIST}
   \mu_{V} = 31.99 \pm 0.13
\end{equation}

Looking back over the Cepheid measurements made in this series 
of papers, we note that the photometry has been most 
troublesome in the those galaxies where the disk is inclined 
to the line of sight. The 2 farthest galaxies of this study, 
NGC~4639 and NGC~3982, presented no photometric inconsistencies. 
Both these galaxies 
present face-on disks. By contrast, it was 
necessary to examine the data most critically for NGC~4527 and 
NGC~3627, which are both closer in distance modulus by at least 
a magnitude. However their disks are very inclined, thus presenting
a higher column density of stars which may be contributing more 
severely to contamination and confusion noise.

%
%

\section{The Type Ia Supernova 1998aq}
SN\,1998aq was discovered in NGC\,3982 by M.~Armstrong (Hurst
et~al. 1998). Pre-maximum and subsequent spectroscopy revealed it to
be a prototype SN\,Ia (Ayani \& Yamaoka 1998; Berlind \& Calkins as
quoted by Garnavich et~al. 1998; Vink{\'o} et~al. 1999). The SN
reached $B$-maximum on JD$\;$2\,450\,931 and had a normal decline 
rate of $\Delta m_{15}=1.12\pm0.05$ (Riess et~al. 1999).

   The light curve of SN\,1998aq, as compiled from amateur
observations by the {\em Variable Star Observers' Network} (VSNET),
indicates an apparent maximum $V$-magnitude of $m_{V}(\max)=12.30
\,(\pm0.15)$. CCD-photometry six days before $B$-maximum showed that
the SN had $m_{V}=12.67 \,(\pm0.05)$ (Hanzl \& Caton 1998). At this
epoch a SN\,Ia is $0.40 \,(\pm0.05)$ below $V$-maximum (Leibundgut
1995 - private communication; Riess et~al. 1999). From this follows 
$m_{V}(\max)=12.27
\,(\pm0.10)$ in fortuitous agreement with the above value. We will
adopt $m_{V}(\max)=12.28\pm0.08$. 

   On the assumption that SN\,1998aq suffers the same reddening of
$E(B-V)=0.09$ (of which 0.01 is due to Galactic reddening;
Schlegel, Finkbeiner, \& Davis 1998) as an average Cepheid in
NGC\,3982, the absolute magnitude of the SN is obtained by simply
subtracting the {\em apparent\/} $V$-modulus of $\mu_{V}=31.99\pm0.13$
from the {\em apparent\/} maximum $V-$magnitude of the SN of 
$m_{V}(\max)=12.28\pm0.08$. This gives 
\begin{equation}
   M_{V}(\max)=-19.71\pm0.15  
\end{equation}
for SN\,1998aq.
%
%
This value is brighter than that of any
of the eight Cepheid-calibrated normal SNe\,Ia (Parodi et~al. 2000,
Tammann, Sandage, \& Saha 2001), but not significantly so.  
It is therefore unlikely that we have
underestimated the absorption of the SN. If the absolute magnitude is
reduced to the standard decline rate of $\Delta m_{15}=1.2$ by means
of equation~(10) by Parodi et~al. (2000) one obtains
\begin{equation}\label{eq:SN1}
   M_{V}^{\rm corr}=-19.67\pm0.15.
\end{equation}

Of course, it is possible that SN\,1998aq suffers only the Galactic
reddening of $E(B-V)=0.014$ and very little or no additional reddening
in its parent galaxy. The absence of Na\,D absorption lines in its
spectrum (Ayani \& Yamaoka 1998) may be taken as a hint in this
direction. If the SN indeed suffered no absorption in NGC\,3982, its
absolute magnitude would become fainter by $3.1\times
0.076=0.24\;$mag than quoted above, i.e.
\begin{equation}
   M_{V}(max) = -19.47 \pm 0.15
\end{equation} 
\begin{equation}\label{eq:SN2}
   M_{V}^{\rm corr}=-19.43\pm0.15.
\end{equation}
This is a firm lower limit to the luminosity of SN\,1998aq because it
cannot suffer less than zero absorption. Yet it is only
insignificantly fainter than the {\em mean\/} of the eight
Cepheid-calibrated SNe\,Ia of $<\!M_{V}^{\rm corr}\!>=-19.48\pm0.07$
(Parodi et~al. 2000).

   Carrying the discussion one step further, it is noted that
SN\,1998aq is quoted by an anonymous referee to the paper by
Vink{\'o} et~al. (1999) 
as having had $(B-V)=-0.17$ six days before $B$-maximum. 
With $m_{V}=12.67$ from above for the same epoch 
and considering that a SN\,Ia is then $0.28\;$mag below $B$-maximum
(Leibundgut 1995 -- private communication; Riess et~al. 1999), one 
obtains $m_{B}(\max)=12.22
\,(\pm0.10)$. From this follows $(B-V)=-0.06 \,(\pm0.14)$, i.e. a value
which happens to agree very well with the mean intrinsic color of
normal SNe\,Ia with $<\!B-V\!>=-0.013$, $\sigma=0.06$ (Parodi
et~al. 2000). It seems therefore that SN\,1998aq indeed suffers very
little absorption in its parent galaxy.

   It should be mentioned that there is another color measurement of
SN\,1998aq six days before $B$-maximum giving $(B-V)=+0.02$ (Hanzl \&
Caton 1998). This very red pre-maximum color would imply a reddening
of $E(B-V)\approx0.2$. The corresponding absorption of
$A_{B}\approx0.8$ and $A_{V}\approx0.6$ would give to SN\,1998aq an
absolute magnitude around $M_{B}(\max)\approx M_{V}(\max)\approx -20.0$.
The only SN\,Ia known to have become so luminous is SN\,1995ac (Saha
et~al. 2001) which, however, had a peculiar SN\,1991T-like spectrum
and which is definitely not shared by SN\,1998aq. We conclude
therefore that the published color of $(B-V)=+0.2$ must be erroneous.

   If the adopted values of $m_{B}(\max)=12.22$ and
$m_{V}(\max)=12.28$ are corrected for Galactic reddening of
$E(B-V)=0.014$ one obtains $m_{B}^{0}(\max)=12.16 \,(\pm0.15)$ and  
$m_{V}^{0}(\max)=12.24 \,(\pm0.14)$. With a true modulus of NGC\,3982 of
$(m-M)^{0}=31.72\pm0.14$ the absolute magnitudes become then
\begin{equation}\label{eq:SN3}
 M_{B}^{0}(\max)=-19.56\pm0.21, ~~~~ M_{V}^{0}(\max)=-19.48\pm0.20.
\end{equation}
These values can now be reduced to the standard values of $\Delta
m_{15}=1.2$ and $(B-V)=-0.01$ using equations~(9) and (10) of Parodi
et~al. (2000) giving
\begin{equation}\label{eq:SN4}
 M_{B}^{\rm corr}=-19.35\pm0.24, ~~~~ M_{V}^{\rm corr}=-19.34\pm0.23.
\end{equation}
The value of $M_{V}^{\rm corr}$ found here compares well with that in
equation~(\ref{eq:SN2}).

\section{The value of \boldmath{$H_0$}}
There are now nine SNe\,Ia with Branch-normal spectra whose Cepheid
distances are known. They are compiled in Table~\ref{tab:Ho1}. The
reader can find the input data and the original sources in Parodi
et~al. (2000).

\placetable{tab:Ho1}

   SN\,1895B is omitted because its $V$-magnitude at maximum is
unreliable. We have also determined a Cepheid distance for SN\,1991T
(Saha et~al. 2001), but this object should not be used as a calibrator
in view of its peculiar spectrum and its poorly known internal
absorption (cf. however Gibson \& Stetson 2000; Gibson \& Brook
2000). The yet incomplete data on SN\,1991T-like SNe suggest that they
form a quite heterogeneous class (Saha et~al. 2001).

   The absolute magnitudes of the nine calibrating SNe\,Ia in columns
(9)\,$-$\,(11) of Table~\ref{tab:Ho1} are corrected for decline rate
$\Delta m_{15}$ {\em and\/} variations of the intrinsic color
($B_{\max}-V_{\max}$).  They are reduced to a standard SN\,Ia with
$\Delta m_{15}=1.2$ and $(B_{\max}-V_{\max})= -0.01$. The relevant
reduction formulae are given by Parodi et~al. (2000; eq.~9\,$-$\,11; for
quite similar reduction formulae cf. Tripp 1998, and Tripp \& Branch
1999).

   The value of $H_0$ is obtained by fitting the weighted
$<\!\!M_{\lambda}^{\rm corr}\!\!>$ of the calibrators to a sample of
more distant, absorption-free Branch-normal SNe\,Ia. We use here the
fiducial sample of the 35 bluest SNe\,Ia as defined by Parodi
et~al. (2000). They suffer  minimum internal absorption, and
their mean color of $(B_{max}-V_{max}) = -0.012\pm0.008$ perfectly matches 
the mean color of the calibrators of $-0.017 \pm 0.015$. 
The agreement in color, which is not affected by
the corrections for $\Delta m_{15}$ and intrinsic variations of
($B_{\max}-V_{\max}$), is decisive for the determination of $H_0$. It
does not necessarily mean that we have found the true intrinsic color
of SNe\,Ia, but it shows that any remaining absorption is the same for
the calibrators and for the fiducial sample, and this is a sufficient
condition for the derivation of reliable distances.

   The 35 SNe\,Ia of the fiducial sample have a velocity restriction of
$1200 < v \la 30\,000\kms$. The lower limit is set to reduce the
effect of peculiar velocities, the upper limit to keep space curvature
effects small, but still to reflect the {\em large-scale\/} value of
$H_0$. Fitting a straight line to the fiducial sample of the form
\begin{equation}\label{eq:Ho1}
   \log v = 0.2\,m_{\lambda} + c_{\lambda}
\end{equation}
is indistinguishable from the case of a flat Universe with
$\Omega_{\rm M}=1$. A presently favored model with $\Omega_{\rm
  M}=0.3, \Omega_{\Lambda}=0.7$ yields a value of $H_0$ which is only
0.8 units higher (cf. Parodi et.~al. 2000). We pursue the $\Omega_{\rm
  M}=1$ case for the sake of simplicity and comparison with external
results.

   A trivial transformation of equation~(\ref{eq:Ho1}) leads to 
\begin{equation}\label{eq:Ho2}
   \log H_0 = 0.2\,M_{\lambda} + 5 + c_{\lambda}\, .
\end{equation}
The constant $c_{\lambda}$ is defined by the fiducial sample to be
$c_{\rm B}=0.676\pm0.004$, $c_{\rm V}=0.673\pm0.004$, and $c_{\rm
  I}=0.616\pm0.004$. Inserting these values together with the
appropriate values of $<\!\!M_{\lambda}^{\rm corr}\!\!>$ from
Table~\ref{tab:Ho1} into equation~(\ref{eq:Ho2}) yields
$H_0(B)=60.5\pm2.0$, $H_0(V)=60.4\pm1.8$, and  $H_0(I)=60.0\pm2.8$,
i.e. almost identical values in the three wave bands. We adopt at this
point $H_0=60.3\pm1.8$ for the case $\Omega_{\rm M}=1$.

   Gibson et~al. (2000) have re-analyzed the {\em HST\/} Cepheid
observations of Saha et al. (1994, 1995, 1996a,b, 1997, 1999) and Tanvir
et~al. (1995) and have obtained distances which are smaller by
$0\fm14$ on average than the moduli shown in Table~\ref{tab:Ho1}. The
largest deviation of $0\fm49$ is suggested for NGC\,5253; however, the
corresponding small modulus of $(m-M)^0=27.61$ would imply an
exceptionally faint tip of the red-giant branch of NGC\,5253 (Saha
et~al. 1995) as well as of NGC\,5128, another member of the Cen\,A
group. 
Also the surface brightness fluctuation method, albeit with its attendant 
uncertainties, gives $(m-M)^0=28.13$ for NGC\,5128 and for NGC\,5102, which
both belong to the same group (Ferrarese et~al. 2000). 
If NGC\,5253 is
excluded, the Gibson et~al. moduli are still systematically smaller by
$0\fm11\pm0.03$ than adopted in Table~\ref{tab:Ho1} for the first
seven entries. We stand by our original numbers, and emphasize that 
the 118 Cepheids {\it in common} between the work of Gibson et~al. and 
Saha et~al. agree within a few $0\fm01$ (Gibson et~al. 2000, Table~3; 
Parodi et~al. 2000).
The differences in distances derived 
by these 2 sets of reductions must therefore lie in the different samples 
of Cepheids chosen. It is easily demonstrated, say for the specific case 
of NGC~5253, that the sample of common Cepheids with Gibson et al.'s 
photometry yields a distance that is 
significantly higher than what is obtained using all of their Cepheids. 
This is an important point for other reasons as well: the distances 
to all galaxies studied by the Mould, Freedman, Kennicutt et~al. (MFK) 
consortium 
{\it except} for the ones that are a re-analysis of the SNe~Ia bearing 
galaxies first studied by the Sandage et~al. consortium, are based on a 
Cepheids found in {\it common} by the DAOPHOT and DoPHOT based procedures.
For this reason, the Gibson et~al. (2000) re-analysis is not at par with 
the distances for other galaxies studied by the MFK consortium. If the 
Cepheids that are found in common by Gibson et~al. (2000) and by us 
(in the previous papers in this series) are used, even with just the 
Gibson et~al. photometry, the resulting distances to the galaxies 
are in general larger than the ones reported by Gibson et~al. (2000), 
by amounts that vary from one galaxy to another, with NGC~5253 being the 
most striking case.
Despite these considerations, if the smaller Cepheid distances of Gibson
et~al. (2000) are taken at face value, excluding only NGC\,5253, the
value of $H_0=60.3$ above is increased to $H_0=63.4$. If these authors 
had applied the metallicity corrections of Cepheid distances as now 
adopted by Freedman et~al. (2001), they would have obtained
$H_{0} = 61.7$, a value quite close to ours, but for different reasons.

   Freedman et~al. (2001) have decreased the Gibson et.~al.\ distances
of the first eight galaxies in Table~\ref{tab:Ho1} by an additional
$0\fm09$ on average. The proposed increase is the net result of two
effects. (1) The authors adopt a metallicity dependence of the
Cepheid distances, which {\em  increases\/} their distances by 
$0\fm06$ on average. (2) They use the
Cepheid PL relation in $V$ and $I$ as derived by Udalski et~al.\ (1999)
from OGLE survey data. The unexpectedly flat slope of their $I$-band
PL relation increases the absorption corrections of the long-period
Cepheids and results in a {\em decrease\/} of the Gibson et~al.\
distances by $0\fm15$. The new $I$-band relation needs
further confirmation.  
The new PL relations imply a very flat color-period
relation of
\begin{equation}\label{eq:Ho3}
   (V-I) = 0.202 \log P + {\rm const}\; ,
\end{equation}
which disagrees with the results discussed by Sandage, Bell \& Tripicco
(1999) that are based on the extensive color data of Dean, Warren \& Cousins
(1978) and Caldwell \& Coulson (1984, 1985). The color-period relation
implied by the multi-color PL relations from Madore \& Freedman (1991), 
as well as from Feast \& Walker (1987) are in agreement with the results 
discussed by Sandage, Bell \& Tripicco (1999), but they all disagree with 
the results of Udalski et~al.\ (1999), whose sample does not include 
variables with $\log P > 1.5$ because, as stated in their paper, 
they are too bright and are saturated in the OGLE images. Yet 
such longer-period Cepheids are effectively the ones to which 
the Freedman et~al. (2001) correction applies. 


If we adopt for the moment the new distances of Freedman et al. (2001;
Table~4) for the first eight galaxies in Table~\ref{tab:Ho1} -- which
reduces the adopted moduli by $0\fm25$ on average -- we find
$H_0=67.7$ instead of $H_0=60.3$ from above.

   There remains the question why we obtain $H_0=67.7$ even if we
adopt the Freedman et~al.\ (2001) distances at face value, while these
authors suggest $H_0=71$ from SNeIa alone. There are mainly three rather 
subtle reasons. (1) Freedman et~al.\ use only six SNe\,Ia as calibrators. 
Their questionable distance for SN\,1972E in NGC\,5253 therefore contributes 
with higher weight. If this object is excluded, their $H_0$
decreases by 2.5 percent. (2) There is a slight color mismatch between
the calibrators and the distant sample as used by Freedman
et~al. After application of their absorption and decline rate
corrections, their six calibrators have a mean color of $(B-V)^{\rm
  corr}= -0.043\pm0.008$, while the mean color of their distant
SNe\,Ia sample is $(B-V)^{\rm corr}= -0.061\pm0.002$, i.e.\ they are
very blue. Obviously the distant sample is overcorrected for internal
absorption. The difference $\Delta E(B-V)=0.018\pm0.008$ translates
into a reduction of the moduli of the distant SNe\,Ia by $\Delta
m_{\rm B}=0.074$, $\Delta m_{\rm V}=0.056$, and $\Delta m_{\rm
  I}=0.033$. If we correct for this mis-match in color, 
their value for $H_0$ is reduced by another 2.6 percent. (3) The
calibrating SNe\,Ia lie in spirals and hence have slower decline rates
than the distant SNe\,Ia which come in all types of galaxies. In the
case of Freedman et~al.\ (2001) the mean difference amounts to
$\delta\Delta m_{15}=0.20$. This value must be multiplied by the
wavelength-dependent slope of the $\Delta m_{15}$-luminosity relation
to allow a fair comparison between calibrators and distant
SNe\,Ia. The authors have adopted the rather steep slope of Hamuy
et~al.\ (1996), yet an enlarged sample of SNe\,Ia gives a
significantly smaller slope (Parodi et~al.\ 2000; equations~3-5). The
calibrating SNe\,Ia have therefore been overcorrected faintwards by
$0\fm06$, $0\fm04$, and $0\fm04$ in $B$, $V$, and $I$,
respectively. This has led to an overestimate of $H_0$ by $\sim\!2$
percent. -- If these three points are allowed for, the value of
$H_0=71$ by Freedman et~al. (2001) comes very close to the value of
$H_0=67.7$, which we have derived above using their new Cepheid
distances.

   Pending a re-reduction of all {\em HST\/} Cepheid observations
entering Table~\ref{tab:Ho1} we leave it to the reader to weight our
Cepheid distances, those by Gibson et~al. (2000), as well as the ones
by Freedman et~al. (2001) based on Udalski et~al.'s (1999) yet unconfirmed 
PL relation. At present we adopt the distances in Table~\ref{tab:Ho1} and
hence $H_0=60.3\pm1.8 (\Omega_{\rm M}=1)$ or $H_0=61.1\pm1.8
(\Omega_{\rm M}=0.3, \Omega_{\Lambda}=0.7)$.

   It should be noticed that remaining systematic error sources tend
to reduce the true value of $H_0$. In particular there is
increasing evidence that the zero point of the PL relation with LMC
at $(m-M)^0=18.50$ should be made brighter by $\sim\!0\fm06$
(cf. Tammann, Sandage, \& Saha 2001).
Moreover,  if we would apply the metallicity
corrections of the Cepheid distances adopted by Freedman et~al. (2001)
the distances would increase by an additional $0\fm06$. Summarizing
the systematic error sources Parodi et~al. (2000) have conservatively
estimated the correction factor of $H_0$ to be $0.96\pm0.08$. This
leads to 
\begin{equation}\label{eq:Ho4}
   H_0 = 58.7 \pm 6.3\, ,
\end{equation}
including systematic errors and valid for a model with $\Omega_{\rm
  M}=0.3, \Omega_{\Lambda}=0.7$.


  We thank the many individuals at STScI who have worked hard behind the 
scenes to make these observations possible. Support for this work was 
provided by NASA through grant \# HST-GO-08100.01A from the Space 
Telescope Science Institute, which is operated by the Association of 
Universities for Research in Astronomy, Inc., under NASA contract 
NAS5-26555. G.A.T. thanks the Swiss National Science Foundation for 
continued support.

\clearpage


%
%
\clearpage
%

\figcaption[]{Mosaic $V$ image of NGC~3982 showing the field imaged 
with the WFPC2. The orientation on the sky is also indicated.
\label{fig1}}

\figcaption[]{Identifications for all the variable stars found. The numbers 
are the same as in Tables 2 to 5.  Each of the four WFPC2 
chips is shown separately. The orientation on the sky is indicated on each 
of the panels.
\label{fig2}}

\clearpage
\epsscale{0.90}

\clearpage
\plotone{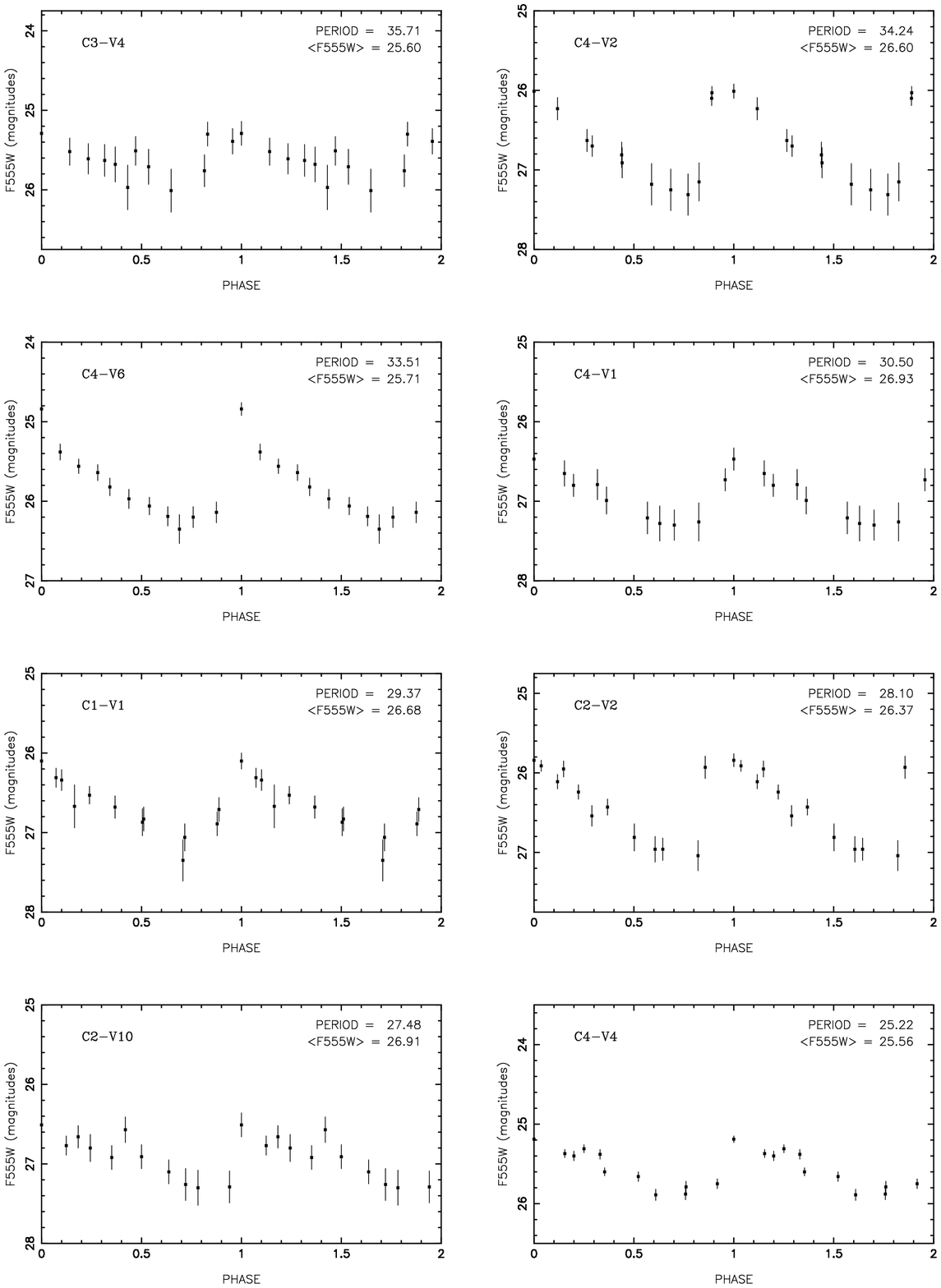}
\clearpage
\plotone{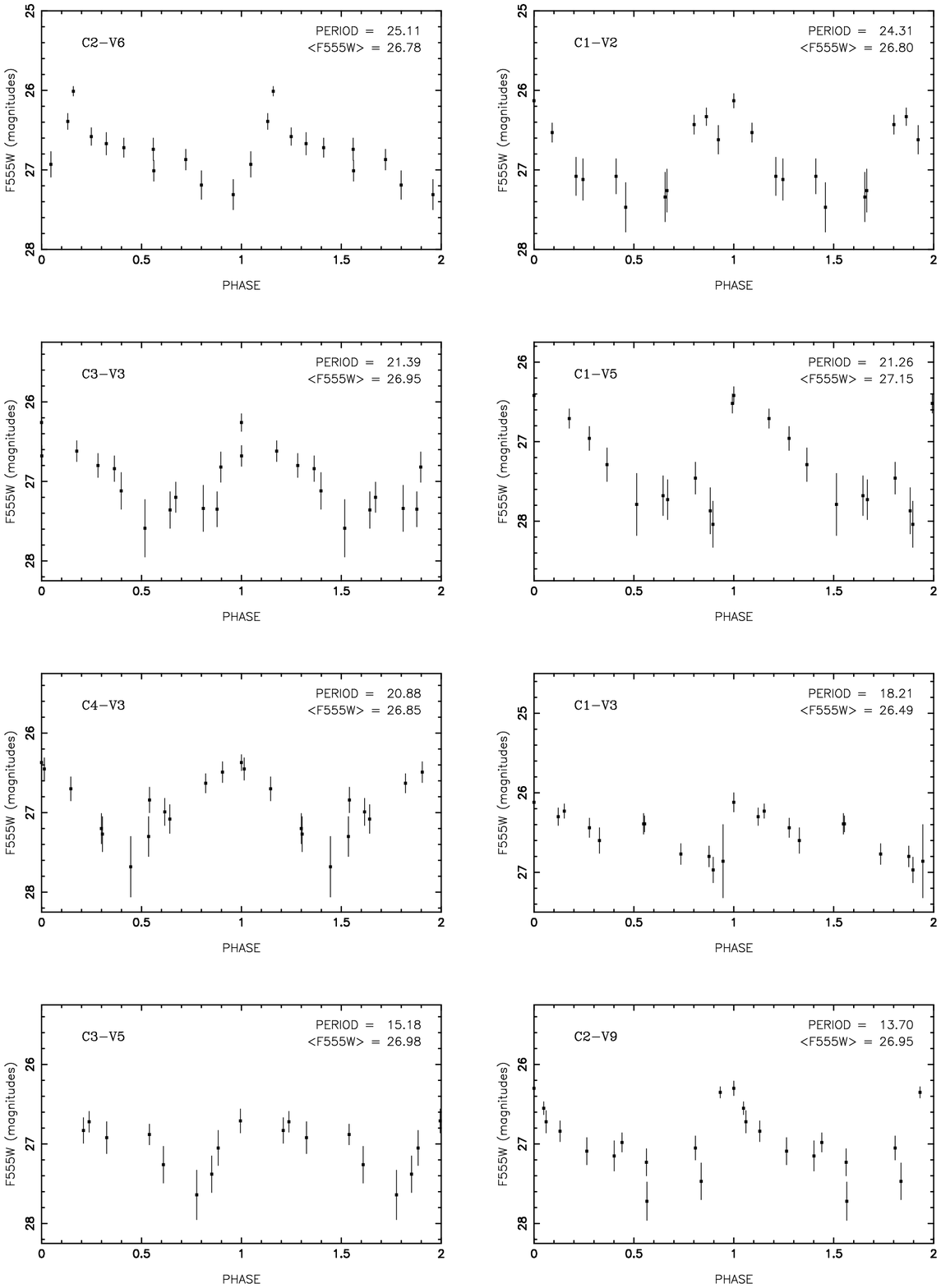}
\clearpage
\plotone{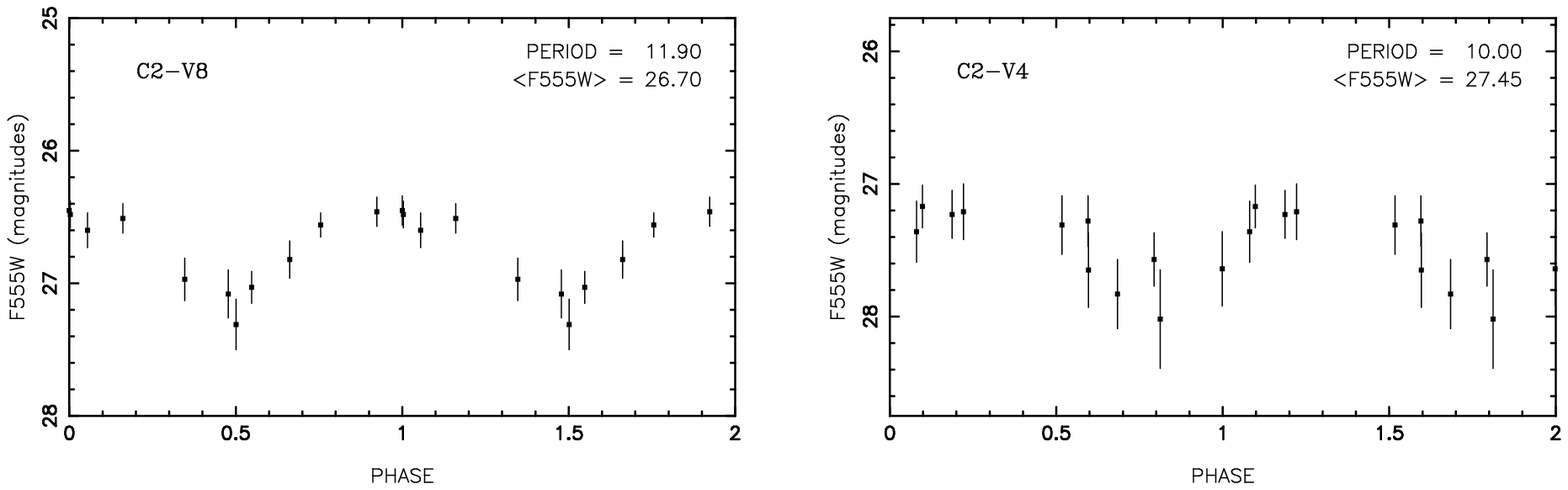}

\clearpage

\clearpage
\plotone{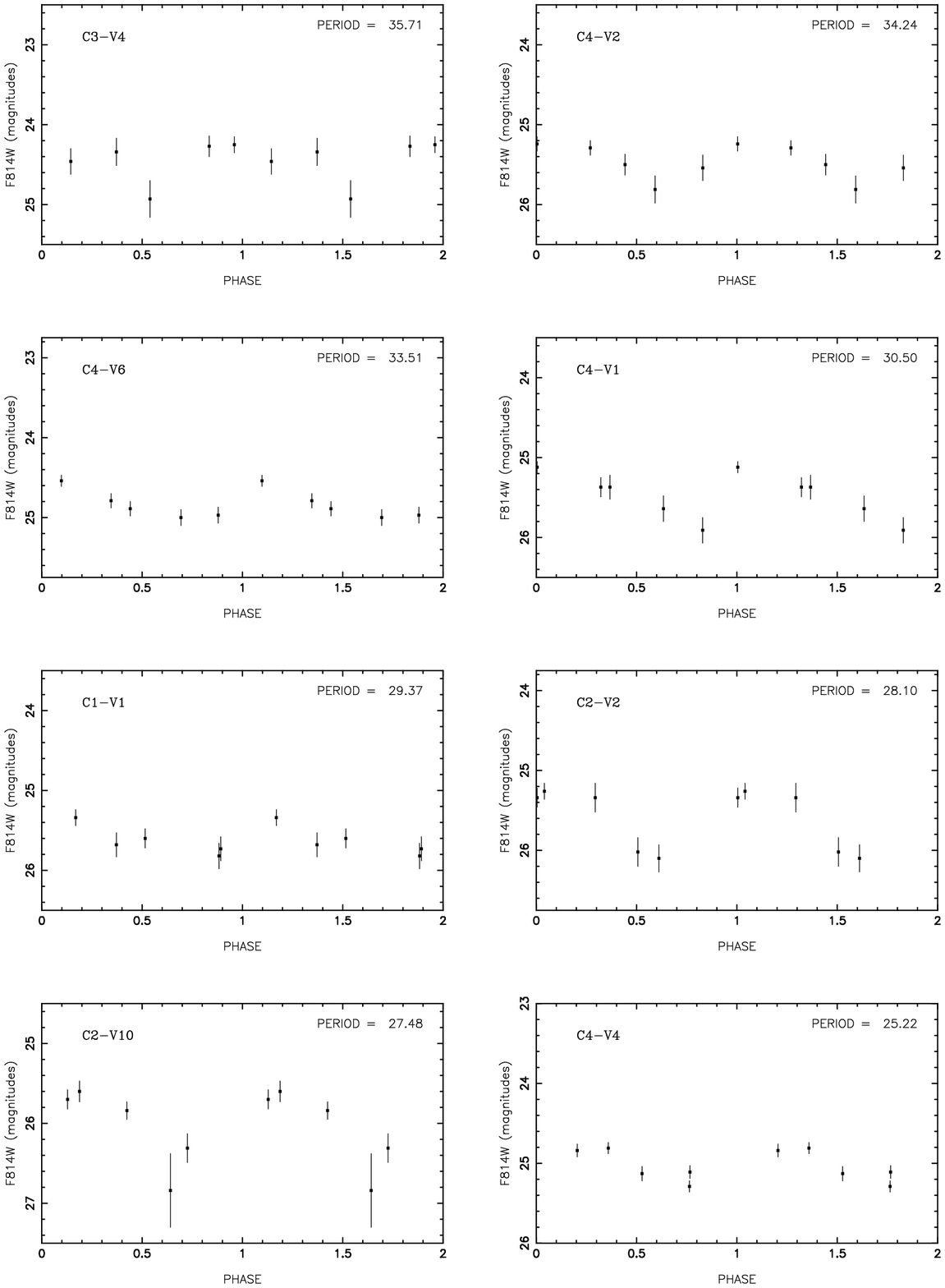}
\clearpage
\plotone{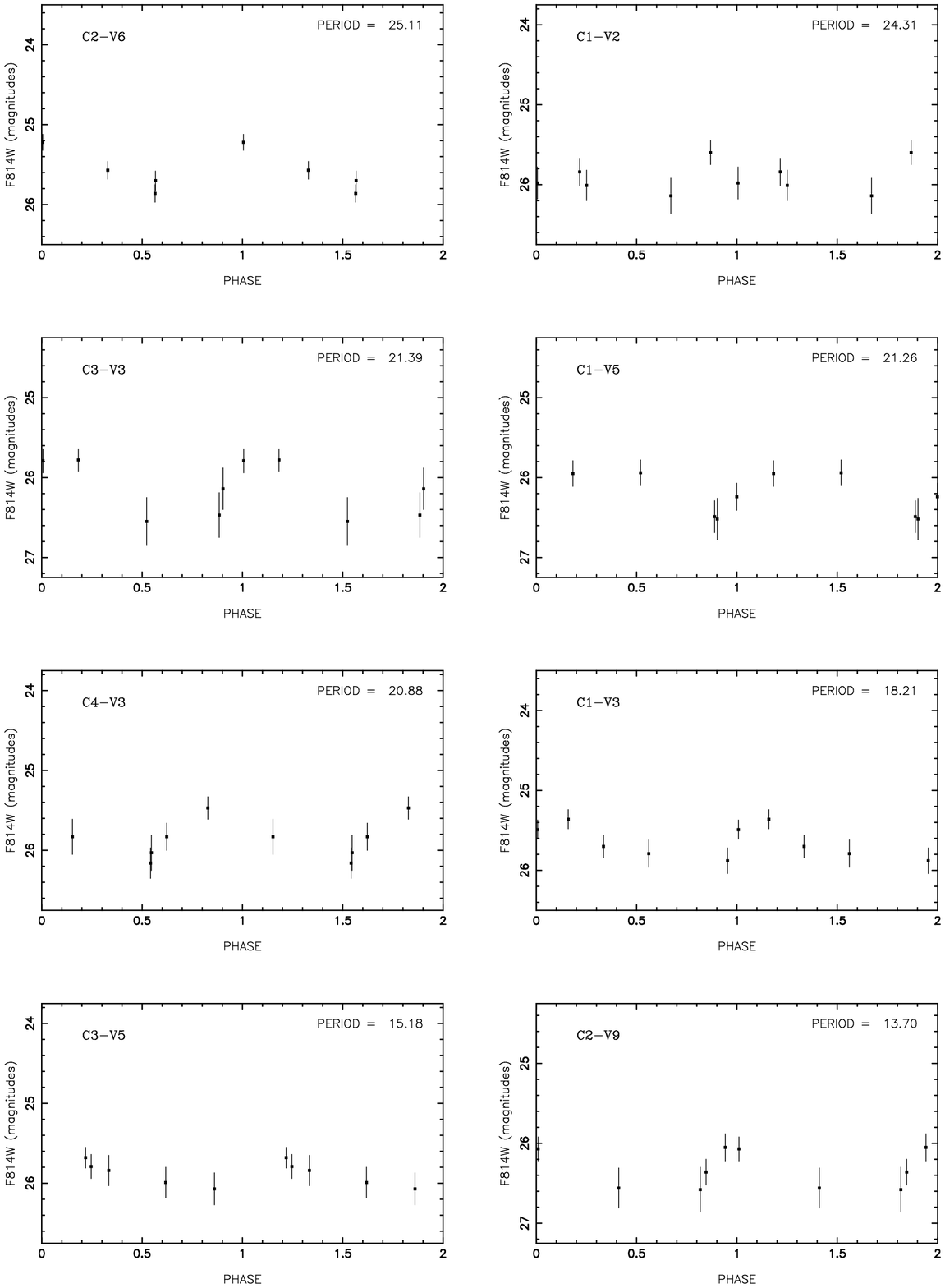}
\clearpage
\plotone{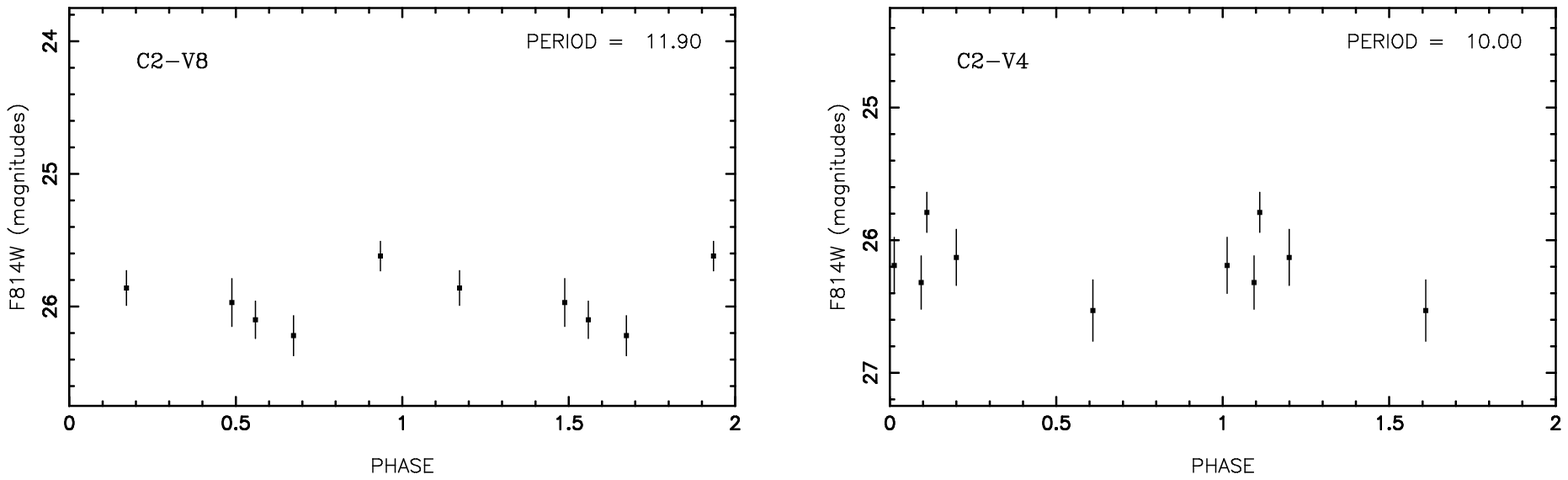}

\clearpage

\clearpage

\clearpage

\clearpage

\clearpage
\setcounter{table}{0}
\begin{deluxetable}{lcc}
\tablecaption{Journal of Observations. \label{tbl1}}
\tablewidth{0pt} 
\tablehead{
\colhead{Data Archive Designation} &\colhead{HJD at Midexposure} &
\colhead{Filter}}
\startdata
u5ky0101r + ...02r & 2451624.018806427 & F555W   \nl
u5ky0201r + ...02r & 2451636.888945493 & F555W   \nl
u5ky0203r + ...04r & 2451637.022626058 & F814W   \nl
u5ky0301r + ...02r & 2451642.923320583 & F555W   \nl
u5ky0401r + ...02r & 2451647.954917874 & F555W   \nl
u5ky0403r + ...04r & 2451648.080265098 & F814W   \nl
u5ky0501r + ...02r & 2451651.255265143 & F555W   \nl
u5ky0601r + ...02r & 2451654.205612403 & F555W   \nl
u5ky0701r + ...02r & 2451656.083042987 & F555W   \nl
u5ky0703r + ...04r & 2451656.214292993 & F814W   \nl
u5ky0801r + ...02r & 2451658.296931908 & F555W   \nl
u5ky0901r + ...02r & 2451662.048668068 & F555W   \nl
u5ky0903r + ...04r & 2451662.180959738 & F814W   \nl
u5ky1001r + ...02r & 2451666.069501458 & F555W   \nl
u5ky1101r + ...02r & 2451672.033737653 & F555W   \nl
u5ky1201r + ...02r & 2451677.060473833 & F555W   \nl 
u5ky1203r + ...04r & 2451677.192418277 & F814W   \nl
\enddata
\tablenotetext{}{All exposures are $2\times2500$s}
\end{deluxetable}
\newpage 
%
%
\begin{deluxetable}{lrr}
\footnotesize
\tablecaption{Position of the Variable Stars on the Deep-V Image
\label{tbl2}}
\tablewidth{0pt}
\tablehead{
\colhead{Variable ID} & \colhead{X-position} & \colhead{Y-position}}
\startdata
C1-V1 & 89.78 & 397.36 \nl
C1-V2 & 129.86 & 108.72 \nl
C1-V3 & 378.66 & 175.47 \nl
C1-V4 & 431.65 & 239.25 \nl
C1-V5 & 600.06 & 260.04 \nl
C2-V1 & 80.19 & 352.03 \nl
C2-V2 & 94.56 & 536.10 \nl
C2-V3 & 120.03 & 511.47 \nl
C2-V4 & 213.90 & 613.04 \nl
C2-V5 & 370.85 & 507.06 \nl
C2-V6 & 458.11 & 399.19 \nl
C2-V7 & 509.04 & 515.58 \nl
C2-V8 & 509.73 & 151.75 \nl
C2-V9 & 542.60 & 524.90 \nl
C2-V10 & 568.24 & 104.44 \nl
C3-V1 & 95.20 & 352.99 \nl
C3-V2 & 297.30 & 601.10 \nl
C3-V3 & 400.07 & 666.87 \nl
C3-V4 & 415.38 & 294.85 \nl
C3-V5 & 447.77 & 597.67 \nl
C4-V1 & 118.46 & 405.86 \nl
C4-V2 & 224.53 & 214.83 \nl
C4-V3 & 267.86 & 232.28 \nl
C4-V4 & 416.15 & 365.64 \nl
C4-V5 & 460.84 & 249.72 \nl
C4-V6 & 530.88 & 134.64 \nl
\enddata
\end{deluxetable}
\newpage
%
%
\begin{deluxetable}{lcccccccc}
\tablecaption{Photometry of Variable Stars: Magnitudes and Error Estimates
\label{tbl3}}
\tiny
\tablewidth{0pt}
\tablehead{ }
\startdata
\tableline\tableline
 HJD & C1-V1 & C1-V2 & C1-V3 & C1-V4 & C1-V5 & C2-V1 & C2-V2 & C2-V3 \nl
\tableline
\multicolumn{9}{c}{$F555W$} \nl
\tableline
 2451624.0188 & 26.31  0.12 &     ---     & 26.60  0.12 &     ---     & 26.21  0.22 & 25.95  0.10 & 25.98  0.12 &     ---      \nl
 2451636.8889 & 26.83  0.16 & 27.08  0.24 & 26.86  0.46 & 27.14  0.16 & 26.52  0.12 & 26.68  0.20 & 26.96  0.16 & 26.14  0.08  \nl
 2451642.9233 & 27.06  0.17 & 27.47  0.31 & 26.44  0.12 & 27.46  0.23 & 26.96  0.15 & 26.57  0.16 & 27.05  0.19 & 26.50  0.13  \nl
 2451647.9549 & 26.71  0.16 & 27.26  0.27 & 26.39  0.10 & 27.26  0.24 & 27.80  0.39 & 25.97  0.10 & 25.84  0.08 & 26.45  0.18  \nl
 2451651.2553 & 26.10  0.11 & 26.43  0.12 & 26.77  0.13 & 27.23  0.17 & 27.73  0.25 & 25.73  0.11 & 26.11  0.09 & 26.05  0.12  \nl
 2451654.2056 & 26.34  0.13 & 26.62  0.18 & 26.97  0.16 & 27.05  0.14 & 27.46  0.20 & 25.98  0.11 & 26.24  0.09 & 25.74  0.08  \nl
 2451656.0830 & 26.67  0.28 & 26.13  0.09 & 26.12  0.12 & 26.84  0.12 & 28.05  0.29 & 25.94  0.10 & 26.55  0.13 & 25.70  0.08  \nl
 2451658.2969 & 26.53  0.12 & 26.53  0.12 & 26.30  0.11 & 26.55  0.10 & 26.42  0.11 & 25.88  0.12 & 26.43  0.10 & 25.63  0.08  \nl
 2451662.0487 & 26.68  0.14 & 27.12  0.26 & 26.60  0.16 & 26.47  0.09 & 26.71  0.12 & 26.01  0.16 & 26.81  0.17 & 25.80  0.08  \nl
 2451666.0695 & 26.88  0.17 & 27.08  0.22 & 26.39  0.13 & 26.73  0.12 & 27.30  0.21 & 26.34  0.13 & 26.96  0.14 & 25.89  0.09  \nl
 2451672.0337 & 27.35  0.26 & 27.34  0.31 & 26.80  0.13 & 26.90  0.13 & 27.68  0.25 & 26.41  0.15 & 25.93  0.14 & 26.05  0.11  \nl
 2451677.0605 & 26.89  0.16 & 26.33  0.11 & 26.23  0.09 & 27.10  0.15 & 27.88  0.29 & 26.52  0.16 & 25.91  0.07 &     ---      \nl
\tableline
\multicolumn{9}{c}{$F814W$} \nl
\tableline
 2451637.0226 & 25.60  0.12 & 25.84  0.17 & 25.88  0.16 & 25.80  0.13 & 26.24  0.17 & 26.16  0.45 & 26.10  0.17 & 25.38  0.16  \nl
 2451648.0803 & 25.73  0.16 & 26.14  0.22 & 25.80  0.17 & 25.99  0.15 & 25.94  0.16 & 25.42  0.16 & 25.34  0.12 & 25.80  0.18  \nl
 2451656.2143 & 25.34  0.11 & 25.98  0.20 & 25.49  0.12 & 25.84  0.13 & 26.52  0.26 & 25.03  0.13 & 25.34  0.18 & 25.23  0.11  \nl
 2451662.1809 & 25.68  0.16 & 26.01  0.19 & 25.70  0.14 & 25.56  0.11 & 25.95  0.16 & 25.24  0.14 & 26.02  0.18 & 25.42  0.14  \nl
 2451677.1924 & 25.82  0.16 & 25.60  0.15 & 25.36  0.12 & 26.02  0.16 & 26.49  0.20 & 25.84  0.28 & 25.26  0.10 & 25.45  0.13  \nl
\tableline\tableline
  HJD & C2-V4 & C2-V5 & C2-V6 & C2-V7 & C2-V8 & C2-V9 & C2-V10 & C3-V1 \nl
\tableline
\multicolumn{9}{c}{$F555W$} \nl
\tableline         
 2451624.0188 & 27.57  0.20 & 27.23  0.24 & 26.93  0.16 & 26.34  0.10 &     ---     & 26.72  0.14 &     ---     & 25.32  0.20  \nl
 2451636.8889 & 27.36  0.23 & 26.42  0.09 & 26.74  0.14 & 26.67  0.11 & 27.03  0.12 & 26.30  0.09 & 27.26  0.20 & 25.03  0.16  \nl
 2451642.9233 & 27.83  0.26 & 26.44  0.10 & 27.19  0.18 & 27.13  0.20 & 26.60  0.13 & 26.98  0.12 & 27.30  0.20 & 25.02  0.14  \nl
 2451647.9549 & 27.23  0.19 & 26.76  0.11 & 25.78  0.18 & 27.41  0.17 & 27.08  0.18 & 27.05  0.15 & 26.77  0.12 & 25.32  0.20  \nl
 2451651.2553 & 27.31  0.22 & 26.92  0.13 & 26.39  0.10 & 27.33  0.20 & 26.56  0.09 & 26.55  0.08 & 26.80  0.17 & 25.17  0.16  \nl
 2451654.2056 & 28.02  0.38 & 26.92  0.18 & 26.58  0.11 & 27.32  0.17 & 26.48  0.10 & 27.09  0.17 & 26.92  0.15 & 24.76  0.15  \nl
 2451656.0830 & 25.27  0.14 & 27.27  0.20 & 26.67  0.14 & 27.05  0.15 & 26.51  0.11 & 27.15  0.19 & 26.57  0.16 & 24.53  0.10  \nl
 2451658.2969 & 27.21  0.21 & 27.22  0.19 & 26.72  0.12 & 27.10  0.16 & 26.97  0.16 & 27.23  0.17 & 26.91  0.15 & 24.66  0.15  \nl
 2451662.0487 & 27.65  0.28 & 27.80  0.29 & 27.01  0.13 & 26.27  0.07 & 26.82  0.14 & 27.47  0.23 & 27.10  0.15 & 24.39  0.09  \nl
 2451666.0695 & 27.64  0.28 & 26.89  0.11 & 26.88  0.13 & 26.30  0.13 & 26.45  0.11 & 26.84  0.13 & 27.30  0.22 & 24.61  0.11  \nl
 2451672.0337 & 27.28  0.19 & 26.19  0.07 & 27.31  0.19 & 26.46  0.10 & 27.31  0.19 & 27.72  0.24 & 26.51  0.15 & 24.90  0.16  \nl
 2451677.0605 & 27.17  0.16 & 26.39  0.08 & 26.01  0.06 & 26.85  0.14 & 26.46  0.11 & 26.35  0.07 & 26.66  0.14 & 24.94  0.14  \nl
\tableline
\multicolumn{9}{c}{$F814W$} \nl
\tableline
 2451637.0226 & 26.32  0.20 & 25.75  0.09 & 25.86  0.11 & 25.45  0.11 & 26.10  0.14 & 26.07  0.15 & 26.31  0.18 & 24.55  0.34  \nl
 2451648.0803 & 26.13  0.21 & 25.56  0.12 & 25.22  0.10 & 25.60  0.12 & 25.97  0.18 & 26.58  0.28 & 25.70  0.12 & 24.64  0.39  \nl
 2451656.2143 & 26.19  0.21 & 26.41  0.23 & 25.57  0.11 & 26.19  0.17 & 25.86  0.13 & 26.56  0.25 & 25.84  0.11 & 23.82  0.15  \nl
 2451662.1809 & 26.53  0.23 & 26.61  0.19 & 25.70  0.12 & 25.46  0.12 & 26.22  0.15 & 26.36  0.16 & 26.84  0.46 & 23.71  0.16  \nl
 2451677.1924 & 25.80  0.16 & 25.56  0.10 & -4.99  0.05 & 25.44  0.10 & 25.62  0.11 & 26.05  0.17 & 25.60  0.13 & 24.55  0.31  \nl
\tableline\tableline
  HJD & C3-V2 & C3-V3 & C3-V4 & C3-V5 & C4-V1 & C4-V2 & C4-V3 & C4-V4 \nl
\tableline
\multicolumn{9}{c}{$F555W$} \nl
\tableline     
 2451624.0188 & 24.68  0.12 & 27.12  0.23 & 25.51  0.18 & 27.64  0.31 &     ---     & 26.10  0.09 & 26.38  0.10 & 25.31  0.05  \nl
 2451636.8889 & 24.81  0.05 & 26.26  0.11 & 25.30  0.15 & 27.26  0.23 & 26.47  0.14 & 26.63  0.14 & 26.99  0.17 & 25.80  0.07  \nl
 2451642.9233 & 24.86  0.05 & 26.80  0.15 & 25.30  0.15 & 26.30  0.14 & 26.80  0.14 & 26.91  0.19 & 26.49  0.13 & 25.19  0.04  \nl
 2451647.9549 & 24.95  0.05 & 27.59  0.36 & 25.52  0.17 & 26.92  0.20 & 26.99  0.17 & 27.18  0.26 & 26.70  0.15 & 25.40  0.06  \nl
 2451651.2553 & 24.88  0.05 & 27.20  0.19 & 25.61  0.19 & 26.88  0.13 &     ---     & 27.25  0.26 & 27.27  0.22 & 25.38  0.06  \nl
 2451654.2056 & 24.98  0.05 & 27.34  0.29 & 25.63  0.20 &     ---     & 27.21  0.20 & 27.31  0.26 & 27.68  0.38 &     ---     \nl 
 2451656.0830 & 24.96  0.05 & 26.82  0.19 & 25.68  0.22 & 27.38  0.23 & 27.28  0.22 & 27.15  0.24 & 27.30  0.25 & 25.66  0.06  \nl
 2451658.2969 & 25.01  0.05 & 26.68  0.13 & 25.97  0.28 & 26.71  0.15 & 27.30  0.19 & 26.03  0.08 & 27.08  0.18 & 25.89  0.07  \nl
 2451662.0487 & 24.92  0.04 & 26.62  0.13 & 25.71  0.22 & 26.72  0.13 & 27.26  0.24 & 26.01  0.09 & 26.63  0.12 & 25.88  0.07  \nl
 2451666.0695 & 24.91  0.05 & 26.84  0.16 & 26.01  0.27 &     ---     & 26.73  0.14 & 26.23  0.14 & 26.45  0.14 & 25.75  0.06  \nl
 2451672.0337 & 24.68  0.04 & 27.36  0.23 & 25.76  0.20 & 27.05  0.22 & 26.65  0.16 & 26.70  0.13 & 27.20  0.19 & 25.38  0.05  \nl
 2451677.0605 & 24.74  0.04 & 27.35  0.22 & 25.39  0.16 & 26.83  0.16 & 26.80  0.19 & 26.81  0.16 & 26.84  0.16 & 25.60  0.05  \nl
\tableline
\multicolumn{9}{c}{$F814W$} \nl
\tableline
 2451637.0226 & 24.74  0.10 & 25.80  0.15 & 24.27  0.13 & 25.99  0.19 & 25.12  0.07 & 25.30  0.09 & 25.83  0.17 & 25.11  0.08  \nl
 2451648.0803 & 24.66  0.09 & 26.55  0.30 & 24.46  0.16 & 25.84  0.19 & 25.38  0.15 & 25.81  0.17 & 25.83  0.22 & 24.84  0.08  \nl
 2451656.2143 & 24.70  0.11 & 26.14  0.26 & 24.34  0.17 & 26.07  0.20 & 25.64  0.16 & 25.55  0.16 & 26.16  0.19 & 25.13  0.09  \nl
 2451662.1809 & 24.69  0.10 & 25.78  0.14 & 24.93  0.23 & 25.80  0.15 & 25.91  0.16 & 25.24  0.09 & 25.47  0.14 & 25.30  0.07  \nl
 2451677.1924 & 24.75  0.09 & 26.47  0.28 & 24.25  0.10 & 25.68  0.13 & 25.38  0.12 & 25.50  0.13 & 26.03  0.22 & 24.81  0.07  \nl
\tableline\tableline     
 HJD &  C4-V5 & C4-V6 &&&&& \nl  
\tableline
\multicolumn{9}{c}{$F555W$} \nl
\tableline   
 2451624.0188 & 26.06  0.09 & 26.66  0.24  \nl
 2451636.8889 & 25.63  0.10 & 25.38  0.10  \nl
 2451642.9233 & 25.62  0.21 & 25.64  0.10  \nl
 2451647.9549 & 25.56  0.10 & 25.97  0.12  \nl
 2451651.2553 & 25.57  0.11 & 26.06  0.11  \nl
 2451654.2056 & 25.93  0.10 & 26.19  0.12  \nl
 2451656.0830 & 25.88  0.10 & 26.35  0.18  \nl
 2451658.2969 & 26.01  0.11 & 26.20  0.13  \nl
 2451662.0487 & 26.75  0.25 & 26.14  0.13  \nl
 2451666.0695 & 26.41  0.22 & 24.84  0.08  \nl
 2451672.0337 & 25.98  0.10 & 25.56  0.09  \nl
 2451677.0605 & 25.91  0.11 & 25.82  0.11  \nl
\tableline
\multicolumn{9}{c}{$F814W$} \nl
\tableline
 2451637.0226 & 25.64  0.16 & 24.55  0.07  \nl
 2451648.0803 &     ---     & 24.89  0.09  \nl
 2451656.2143 & 25.59  0.14 & 25.00  0.10  \nl
 2451662.1809 &     ---     & 24.97  0.10  \nl
 2451677.1924 & 25.51  0.12 & 24.80  0.09  \nl
\enddata
\end{deluxetable}
\newpage
%
%
\begin{deluxetable}{lcccccccrcc}	
\tablecaption{Characteristics of the Cepheids \label{tbl4}}	
\scriptsize	
\tablewidth{0pt}	
\tablehead{	
 \colhead{Object} & \colhead{Period} & \colhead{$\langle{V}\rangle$} &	
 \colhead{$\sigma_{\langle{V}\rangle}$} & \colhead{$\langle{I}\rangle$} & 	
 \colhead{$\sigma_{\langle{I}\rangle}$} & \colhead{$U_{V}$} & 	
 \colhead{$U_{I}$} & \colhead{$U_{T}$} & \colhead{$\sigma_{U_{T}}$} & 	
 \colhead{Quality} \\	
 \colhead{} & \colhead{(days)} & \colhead{} & 	
 \colhead{} & \colhead{} & \colhead{} &	
 \colhead{} & \colhead{} & \colhead{} &	
 \colhead{} & \colhead{Index} \\	
 \colhead{(1)} & \colhead{(2)} & \colhead{(3)} &	
 \colhead{(4)} & \colhead{(5)} & \colhead{(6)} &	
 \colhead{(7)} & \colhead{(8)} & \colhead{(9)} &	
 \colhead{(10)} & \colhead{(11)}	
 } 	
\startdata	
C1-V1 & 29.37 & 26.71 & 0.17 & 25.63 & 0.16 & 32.16 & 31.93 & 31.61 & 0.51 & 5 \nl 
C1-V2 & 24.31 & 26.83 & 0.22 & 25.94 & 0.24 & 32.06 & 31.99 & 31.88 & 0.71 & 3 \nl 
C1-V3 & 18.21 & 26.52 & 0.16 & 25.65 & 0.26 & 31.40 & 31.31 & 31.19 & 0.72 & 4 \nl 
C1-V4 & 43.00 & 26.97 & 0.16 & 25.83 & 0.21 & 32.88 & 32.64 & 32.28 & 0.61 & 4 \nl 
C1-V5 & 21.26 & 27.19 & 0.24 & 26.08 & 0.32 & 32.25 & 31.95 & 31.53 & 0.88 & 3 \nl 
C2-V1 & 48.60 & 26.27 & 0.15 & 25.40 & 0.37 & 32.32 & 32.37 & 32.44 & 0.97 & 5 \nl 
C2-V2 & 28.10 & 26.40 & 0.13 & 25.67 & 0.17 & 31.80 & 31.91 & 32.07 & 0.53 & 5 \nl 
C2-V3 & 40.50 & 26.06 & 0.11 & 25.53 & 0.12 & 31.89 & 32.25 & 32.76 & 0.42 & 3 \nl 
C2-V4 & 10.00 & 27.49 & 0.25 & 26.17 & 0.32 & 31.65 & 31.04 & 30.18 & 0.89 & 0 \nl
C2-V5 & 37.30 & 26.76 & 0.15 & 25.89 & 0.38 & 32.49 & 32.51 & 32.53 & 0.99 & 4 \nl
C2-V6 & 25.11 & 26.82 & 0.14 & 25.60 & 0.63 & 32.08 & 31.69 & 31.14 & 1.56 & 5 \nl 
C2-V7 & 40.10 & 26.81 & 0.15 & 25.60 & 0.29 & 32.63 & 32.32 & 31.86 & 0.79 & 1 \nl 
C2-V8 & 11.90 & 26.74 & 0.13 & 25.94 & 0.20 & 31.10 & 31.04 & 30.95 & 0.57 & 4 \nl 
C2-V9 & 13.70 & 26.99 & 0.16 & 26.28 & 0.26 & 31.52 & 31.57 & 31.64 & 0.73 & 4 \nl 
C2-V10 & 27.48 & 26.94 & 0.17 & 25.98 & 0.44 & 32.31 & 32.19 & 32.03 & 1.13 & 3 \nl 
C3-V1 & 41.03 & 24.90 & 0.15 & 24.06 & 0.46 & 30.75 & 30.80 & 30.88 & 1.17 & 5 \nl 
C3-V2 & 42.32 & 24.89 & 0.06 & 24.84 & 0.17 & 30.78 & 31.63 & 32.83 & 0.49 & 1 \nl 
C3-V3 & 21.39 & 26.99 & 0.22 & 26.09 & 0.34 & 32.06 & 31.98 & 31.85 & 0.92 & 5 \nl 
C3-V4 & 35.71 & 25.64 & 0.20 & 24.54 & 0.23 & 31.32 & 31.10 & 30.78 & 0.68 & 0 \nl 
C3-V5 & 15.18 & 27.01 & 0.20 & 25.91 & 0.14 & 31.67 & 31.34 & 30.85 & 0.50 & 3 \nl 
C4-V1 & 30.50 & 26.97 & 0.18 & 25.48 & 0.11 & 32.47 & 31.84 & 30.94 & 0.46 & 5 \nl 
C4-V2 & 34.24 & 26.63 & 0.19 & 25.46 & 0.26 & 32.27 & 31.97 & 31.54 & 0.73 & 5 \nl 
C4-V3 & 20.88 & 26.88 & 0.20 & 25.72 & 0.18 & 31.93 & 31.57 & 31.06 & 0.59 & 3 \nl 
C4-V4 & 25.22 & 25.59 & 0.06 & 25.01 & 0.14 & 30.86 & 31.11 & 31.47 & 0.42 & 3 \nl 
C4-V5 & 44.43 & 25.93 & 0.15 & 25.75 & 0.15 & 31.87 & 32.60 & 33.65 & 0.49 & 1 \nl 
C4-V6 & 33.51 & 25.74 & 0.12 & 24.79 & 0.17 & 31.35 & 31.26 & 31.14 & 0.51 & 5 \nl 
\enddata
\end{deluxetable}
\newpage
%
%
\begin{deluxetable}{lcccccccrcc}
\tablecaption{HSTphot results for Variables found in Common \label{tbl5}}
\scriptsize
\tablewidth{0pt}
\tablehead{
 \colhead{Object} & \colhead{Period} & \colhead{$\langle{V}\rangle$} &
 \colhead{$\sigma_{\langle{V}\rangle}$} & \colhead{$\langle{I}\rangle$} & 
 \colhead{$\sigma_{\langle{I}\rangle}$} & 
 \colhead{Quality} \\
 \colhead{} & \colhead{(days)} & \colhead{} & 
 \colhead{} & \colhead{} & \colhead{} &
 \colhead{Index} \\
 \colhead{(1)} & \colhead{(2)} & \colhead{(3)} &
 \colhead{(4)} & \colhead{(5)} & \colhead{(6)} &
 \colhead{(7)}
} 
\startdata
C1-V1 & 31.05 & 26.49 & 0.13 & 25.44 & 0.15 & 4 \nl
C1-V2 & 24.90 & 26.59 & 0.14 & 26.16 & 0.18 & 4 \nl
C1-V4 & 37.73 & 26.77 & 0.14 & 25.72 & 0.16 & 3 \nl
C1-V5 & 21.83 & 27.01 & 0.15 & 25.99 & 0.18 & 3 \nl
C2-V2 & 28.10 & 26.53 & 0.19 & 25.80 & 0.26 & 3 \nl
C2-V5 & 37.34 & 26.72 & 0.11 & 25.76 & 0.14 & 4 \nl
C2-V6 & 27.96 & 26.76 & 0.16 & 25.55 & 0.18 & 4 \nl
C2-V7 & 37.14 & 26.76 & 0.15 & 25.75 & 0.18 & 4 \nl
C2-V9 & 13.81 & 26.93 & 0.12 & 26.43 & 0.18 & 2 \nl
C2-V10 & 25.97 & 27.11 & 0.17 & 26.11 & 0.20 & 3 \nl
C3-V3 & 22.65 & 26.89 & 0.22 & 25.96 & 0.24 & 2 \nl
C3-V4 & 32.22 & 25.28 & 0.09 & 24.41 & 0.12 & 1 \nl
C4-V1 & 35.80 & 26.55 & 0.13 & 25.37 & 0.16 & 4 \nl
\enddata
\end{deluxetable}
\newpage
%
%
%
\begin{deluxetable}{llcccccrccc}
\tiny
\tablecaption{Mean absolute $B$, $V$, and $I$ magnitudes of nine
  SNe\,Ia without and with corrections for decline rate and
  color\label{tab:Ho1}} 
\tablewidth{-5pt}
\tablehead{
\colhead{SN} & \colhead{Galaxy} & \colhead{$(m-M)^0$} &
\colhead{$M_{B}^0$} & \colhead{$M_{V}^0$} & \colhead{$M_{I}^0$} &
\colhead{$\Delta m_{15}$} & \colhead{$(B-V)^0$} &
\colhead{$M_{B}^{\rm corr}$} & \colhead{$M_{V}^{\rm corr}$} &
\colhead{$M_{I}^{\rm corr}$} \nl
\colhead{(1)} & \colhead{(2)} & \colhead{(3)} & 
\colhead{(4)} & \colhead{(5)} & \colhead{(6)} & 
\colhead{(7)} & \colhead{(8)} & \colhead{(9)} & 
\colhead{(10)} & \colhead{(11)}
}
\startdata
1937C  & IC\,4182   & 28.36\,(12) & -19.56\,(15) & -19.54\,(17) &    \dots     &
 0.87\,(10) & -0.02 & -19.39\,(18) & -19.37\,(17) &     \dots    \nl
1960F  & NGC\,4496A & 31.03\,(10) & -19.56\,(18) & -19.62\,(22) &    \dots     &
 1.06\,(12) &  0.06 & -19.67\,(18) & -19.65\,(22) &     \dots    \nl
1972E  & NGC\,5253  & 28.00\,(07) & -19.64\,(16) & -19.61\,(17) & -19.27\,(20) &
 0.87\,(10) & -0.03 & -19.44\,(16) & -19.42\,(17) & -19.12\,(20) \nl
1974G  & NGC\,4414  & 31.46\,(17) & -19.67\,(34) & -19.69\,(27) &    \dots     &
 1.11\,(06) &  0.02 & -19.70\,(34) & -19.69\,(27) &     \dots    \nl
1981B  & NGC\,4536  & 31.10\,(12) & -19.50\,(18) & -19.50\,(16) &    \dots     &
 1.10\,(07) &  0.00 & -19.48\,(18) & -19.46\,(16) &     \dots    \nl
1989B  & NGC\,3627  & 30.22\,(12) & -19.47\,(18) & -19.42\,(16) & -19.21\,(14) &
 1.31\,(07) & -0.05 & -19.42\,(18) & -19.41\,(16) & -19.20\,(14) \nl
1990N  & NGC\,4639  & 32.03\,(22) & -19.39\,(26) & -19.41\,(24) & -19.14\,(23) &
 1.05\,(05) &  0.02 & -19.39\,(26) & -19.38\,(24) & -19.02\,(23) \nl
1998bu & NGC\,3368  & 30.37\,(16) & -19.76\,(31) & -19.69\,(26) & -19.43\,(21) &
 1.08\,(05) & -0.07 & -19.56\,(31) & -19.55\,(36) & -19.31\,(21) \nl
1998aq & NGC\,3982  & 31.72\,(14) & -19.56\,(21) & -19.48\,(20) &    \dots     &
 1.12\,(03) & -0.08 & -19.35\,(24) & -19.34\,(23) &     \dots    \nl
\noalign{\smallskip}
\hline
\noalign{\smallskip}
 & \multicolumn{2}{r}{straight mean:} & -19.57\,(04) & -19.55\,(04) & -19.26\,(0
6) &        &       & -19.49\,(04) & -19.47\,(04) & -19.16\,(06) \nl
 & \multicolumn{2}{r}{weighted mean:} & -19.56\,(07) & -19.53\,(06) & -19.25\,(0
9) &        &       & -19.47\,(07) & -19.46\,(06) & -19.19\,(09) \nl
\enddata
\end{deluxetable}

\end{document}